\documentclass[twocolumn, apj, numberedappendix, iop]{openjournal}
\usepackage{natbib}
\usepackage{graphicx,amsmath,amssymb,amstext}
\usepackage{amsbsy,amsfonts,amsthm,color}
\usepackage[colorlinks,linkcolor=blue,citecolor=blue,urlcolor=blue]{hyperref}
\usepackage[utf8]{inputenc}
\usepackage{float}
\usepackage{orcidlink}
\usepackage{booktabs}
\usepackage{xspace}
\usepackage{xcolor}
\usepackage{multirow}

\definecolor{darkgreen}{rgb}{0.0,0.45,0.0}

\newcommand{\galacticus}{\textsc{Galacticus}\xspace}
\newcommand{\romanspace}{\textit{Roman}\xspace}
\newcommand{\euclid}{\textit{Euclid}\xspace}
\newcommand{\emcee}{\texttt{emcee}}
\newcommand{\ha}{H$\alpha$}
\newcommand{\oii}{[O\,{\sc ii}]}
\newcommand{\oiii}{[O\,{\sc iii}]}

\newcommand{\Hamath}{\mathrm{H}\alpha}

\def\mpc{\mathrm{\, Mpc}}
\def\msun{\mathrm{\, M_\odot}}

\defcitealias{2010MNRAS.409..421G}{GB10}
\newcommand{\gbten}{\citetalias{2010MNRAS.409..421G}\xspace}

\begin{document}

\title{Emulator-Assisted Calibration of a Semi-Analytic Galaxy Formation Model for the Roman Galaxy Redshift Survey\vspace{-4ex}}
% A Calibrated \galacticus{} Galaxy Population Model for Roman Grism Survey Simulations
% A Roman-Motivated Calibration of \galacticus{} for Emission-Line Galaxy Mock Catalogues
% Emulator-Assisted Calibration of a Semi-Analytic Galaxy Formation Model for Roman Grism Mock Catalogues
% Emulator-Assisted Calibration of \galacticus{} for Roman Grism Mock Catalogues
% Emulator-Assisted Calibration of a Semi-Analytic Galaxy Formation Model for the Roman Galaxy Redshift Survey

\author{Andrew Robertson$^{*}$\orcidlink{0000-0002-0086-0524}}
\author{Andrew Benson\orcidlink{0000-0001-5501-6008}}
\email{$^*$email: arobertson@carnegiescience.edu}
\affiliation{The Observatories of the Carnegie Institution for Science, 813 Santa Barbara Street, Pasadena, 91101, CA, USA}

\begin{abstract}
Mock catalogues for the \romanspace{} Galaxy Redshift Survey benefit from galaxy
populations whose physical properties and emission-line observables are predicted
within a single galaxy-formation framework. We present an emulator-assisted
calibration of the semi-analytic model \galacticus{} for this purpose. Because
direct posterior exploration with \galacticus{} is computationally prohibitive,
we train Gaussian-process emulators to predict the observables entering the
calibration likelihood. We calibrate the model simultaneously to stellar mass
functions, the star-formation-rate function, size--mass relations, the
black-hole mass--velocity-dispersion relation, the mass--metallicity relation,
and \ha{} luminosity functions. Combining these observables breaks parameter
degeneracies present in individual fits and can reveal tensions where different
observables favour different regions of parameter space. The resulting model
provides a broadly successful simultaneous description of the calibration data,
including the abundance and redshift evolution of \ha{} emitters relevant to the
\romanspace{} Galaxy Redshift Survey. A direct \galacticus{} calculation at the
maximum a posteriori point gives similarly good agreement with the calibration
data, providing a calibrated galaxy-formation model suitable for constructing
physically consistent mock catalogues for  \romanspace{}.
\end{abstract}

%\keywords{}

\maketitle

\section{Introduction}

The Nancy Grace Roman Space Telescope will carry out a wide-field near-infrared grism survey over cosmological volumes, providing galaxy redshifts for large-scale structure measurements and a large sample of emission-line galaxies for studies of galaxy evolution \citep{2015arXiv150303757S, 2022ApJ...928....1W}.
These measurements will extend galaxy redshift surveys to large samples at redshifts where the expansion history and growth of structure are less tightly constrained by existing galaxy surveys.
This is especially timely given recent DESI baryon-acoustic-oscillation (BAO) measurements.
DESI first-year BAO data are consistent with $\Lambda$CDM on their own, but when combined with CMB and Type Ia supernova data they show a mild preference for time evolution of the dark energy equation of state \citep{2025JCAP...02..021D}.
This is not yet conclusive evidence for evolving dark energy, but it does sharpen the case for independent measurements of the distance--redshift relation and growth history.
\romanspace{}  is valuable in this context because it will probe these quantities at high redshift, with a different instrument, galaxy selection, and set of observational systematics from existing ground-based spectroscopic surveys.

Unlike targeted spectroscopic surveys such as DESI \citep{2016arXiv161100036D}, \romanspace{}  will use slitless spectroscopy, dispersing the light from the full field of view.
This is a powerful feature of the survey, because it avoids a hard preselection of targets and can in principle provide spectra for all sufficiently bright sources in the field.
It also makes the data more complicated to interpret.
Because the spectra are dispersed images of the galaxies, spatial and spectral information are mixed together, spectra from neighbouring objects can overlap, and the detectability of an emission line depends not only on its total flux but also on the size and morphology of the galaxy.

The redshifts used for the \romanspace{}  grism survey will be obtained primarily from strong rest-frame optical emission lines, especially \ha{} and \oiii, with \oii{} also relevant at higher redshift.
For \romanspace{}  mock catalogues, the line luminosities must be connected to the underlying galaxy population: stellar masses, star formation rates, gas metallicities, dust attenuation, galaxy sizes, and host halo environments.
These correlations matter both for cosmological analyses, where the number density and bias of selected tracers are important, and for image-level simulations, where source sizes, line fluxes, continuum emission, and blending determine whether a redshift can be recovered.

There are several possible ways to build such mocks. Empirical models such as
\textsc{UniverseMachine} \citep{2019MNRAS.488.3143B} provide powerful
frameworks for connecting galaxies to dark matter haloes, while newer
approaches such as \textsc{Diffsky}\footnote{\href{https://diffsky.readthedocs.io/en/latest/}{https://diffsky.readthedocs.io/en/latest/}} extend this framework to star-formation
histories and spectral-energy-distribution modelling \citep{2023MNRAS.518..562A, 2023MNRAS.521.1741H}. Semi-analytic models
offer a complementary approach, evolving galaxies along dark matter halo
merger trees using approximate, physically motivated prescriptions for
baryonic processes \citep[e.g.][]{2006RPPh...69.3101B,2015ARA&A..53...51S}.
They are far less computationally expensive than hydrodynamical simulations,
but still explicitly model processes including gas cooling, star formation,
feedback, metal enrichment, galaxy sizes, black-hole growth, and nebular
emission. For \romanspace{}  mock catalogues, this provides a natural way to predict
many correlated properties of each galaxy within a single model.

The flexibility of semi-analytic models also makes their calibration a
non-trivial part of the modelling problem. Their physical prescriptions
contain parameters that are only weakly constrained by first-principles
arguments, and different combinations of parameters can produce similar
agreement with a small set of observables. Automated calibration methods have therefore become increasingly important.
These include methods that sample the posterior distribution of model
parameters \citep[e.g.][]{2009MNRAS.396..535H,2011MNRAS.416.1949L,
2015MNRAS.451.2663H}, as well as optimisation methods designed to search
efficiently for well-fitting regions of parameter space
\citep[e.g.][]{2015ApJ...801..139R}. Comparisons in which multiple
semi-analytic models are calibrated against a common set of observations
have also provided insight into the successes and limitations of different
models \citep[e.g.][]{2018MNRAS.475.2936K}.

Although semi-analytic models are computationally inexpensive compared with
hydrodynamical simulations, evaluating them directly for the large number of
parameter combinations required for posterior inference remains prohibitively
expensive. One solution is to construct an \emph{emulator}: a statistical
model trained on a finite set of direct model evaluations that predicts the
model output at new points in parameter space at much lower computational
cost. Several studies have used such emulators to explore or calibrate
semi-analytic models. For \textsc{Galform}, \citet{2010MNRAS.407.2017B}
combined emulation with so-called \emph{history matching} to identify the
region of a 16-dimensional parameter space compatible with local luminosity
functions, an approach later applied to the evolving stellar mass function
by \citet{2017MNRAS.466.2418R}. More recently,
\citet{2021MNRAS.506.4011E} used deep-learning emulators to calibrate
\textsc{Galform} to a broad set of low-redshift observables, while
\citet{2024MNRAS.535.3324M} applied a related approach to \ha{} emitters for
\euclid{} and \romanspace{} survey predictions.

Our previous calibration of \galacticus{} used a different strategy to reduce
the computational cost \citep{2026OJAp....955306R}. Rather than evolving a
large ensemble of merger trees for every likelihood evaluation, we calibrated
the model using small samples of galaxies at a few fixed halo masses. This
approach is particularly well suited to quantities for which there are
observational constraints at fixed halo mass, with stellar mass being the
clearest example
\citep[e.g. observationally inferred stellar-to-halo mass relations
from][]{2012ApJ...744..159L,2019MNRAS.488.3143B}. The resulting calibration
produced a good match to the observed stellar mass function. However, matching stellar mass functions at several redshifts primarily
constrains the total amount of star formation between those redshifts, and
does not ensure that this star formation is distributed correctly among
galaxies. The global star-formation history can therefore be approximately
correct while the distribution of galaxy star-formation rates is not.
Consequently, the model in \citet{2026OJAp....955306R} reproduced stellar
mass functions more successfully than \ha{} luminosity functions.

For the \romanspace{}  grism survey, this distinction is important because the redshift
sample is built from emission-line detections. In this work we therefore use
full \galacticus{} runs to calculate the observables entering the likelihood,
including emission-line luminosity functions, and make the resulting posterior
inference computationally feasible using Gaussian-process emulators. We
calibrate simultaneously to a broad set of observables that probe different
aspects of the galaxy formation model, including stellar masses, star-formation
rates, galaxy sizes, metallicities, black-hole masses, and \ha{} luminosities.
We focus primarily on low- and intermediate-redshift calibration data, where
the observational constraints are generally better established, while also
including \ha{} luminosity functions at higher redshift because of their direct
relevance to the \romanspace{}  grism survey.

The emulators are trained on a finite set of full \galacticus{} calculations
and provide both predictions and associated emulator uncertainties, which we take into account when comparing the model predictions with observational data. We test their accuracy using
held-out \galacticus{} runs, use them to sample the posterior distribution of
the model parameters, and finally verify the resulting calibrated model with a direct \galacticus{} calculation. The resulting calibration is intended both
for \romanspace{}  grism mock generation and for broader applications requiring
physically motivated, self-consistent galaxy catalogues.

The paper is organised as follows.
Section~\ref{sec:model} describes the \galacticus{} model, the varied parameters, and the training campaign.
Section~\ref{sec:emulator} describes the emulator construction and validation.
Section~\ref{sec:posterior-inference} describes the likelihood and MCMC inference.
Section~\ref{sec:calibration-results} presents the calibration results, including diagnostic subset fits and the final joint calibration.
Section~\ref{sec:discussion} discusses the interpretation and limitations of the calibration.
We summarise in Section~\ref{sec:conclusions}.

\section{Model And Emulator Training Set}
\label{sec:model}

\subsection{The \galacticus{} Model}

We use \galacticus{}, a semi-analytic model of galaxy formation \citep{2012NewA...17..175B}.
For a given dark matter halo accretion and merger history, encoded in the so-called \emph{merger tree} \citep{1993MNRAS.262..627L}, \galacticus{} follows the formation and evolution of the galaxies associated with each halo and subhalo.
The calculation includes parameterised prescriptions for gas cooling, star formation, stellar feedback, AGN feedback, galaxy sizes, galaxy mergers, black-hole growth, and metal enrichment.

The fixed modelling choices used here are mostly the same as in
\citet{2026OJAp....955306R}, while we vary a substantially larger set of
numerical parameters within this model, as described in
Section~\ref{sec:varied-parameters}. We therefore focus here on the fixed
modelling choices that differ from \citet{2026OJAp....955306R} and on the
parameters varied during the calibration.

The \galacticus{} code is open source,\footnote{\href{https://github.com/galacticusorg/galacticus}{https://github.com/galacticusorg/galacticus}} with detailed documentation of its physical prescriptions available online.\footnote{\href{https://galacticus.readthedocs.io/en/stable/index.html}{https://galacticus.readthedocs.io/en/stable/index.html}}
The emulator and calibration code used in this work is publicly available in
the \textsc{GalacticEmu} repository.\footnote{\href{https://github.com/Andrew-Robertson/Galacticus-emulation}{https://github.com/Andrew-Robertson/Galacticus-emulation}}
The complete \galacticus{} parameter files and the larger data products needed
to reproduce the analysis will be deposited in a versioned Zenodo repository
upon acceptance of the paper.

The main fixed model choice that differs from \citet{2026OJAp....955306R} is the treatment of galaxy sizes.
In that work, galaxy component sizes were computed using an iterative equilibrium calculation, in which the baryons contribute to the gravitational potential used to determine component radii.
In the larger-volume runs needed here, this produced a small sub-population of galaxies with unrealistically small sizes. %scale lengths roughly 10--100 times smaller than typical galaxies of the same stellar mass.
We therefore use a simpler size calculation, in which component radii are determined from their specific angular momentum in the dark matter halo potential, without allowing the baryons to modify that potential.\footnote{Specifically, \citet{2026OJAp....955306R} used
\href{https://galacticus.readthedocs.io/en/v0.9.12/physics/galacticStructureSolver.html\#galacticstructuresolverequilibrium}{\texttt{galacticStructureSolverEquilibrium}},
whereas here we use
\href{https://galacticus.readthedocs.io/en/v0.9.12/physics/galacticStructureSolver.html\#galacticstructuresolversimple}{\texttt{galacticStructureSolverSimple}}.}
%The origin of the small-size tail is not yet fully understood, but one possibility is that the \emph{pseudo-angular-momentum} treatment used for spheroid sizes is being applied outside the regime where it is physically appropriate, given that these systems are not typically rotationally supported.

\subsection{Merger Trees And Halo Sampling}
\label{sec:tree-sampling}

For the training runs, \galacticus{} is run on extended Press--Schechter
merger trees following the
method of \citet{2008MNRAS.383..557P}. We sample 17,385 $z=0$ host haloes over the mass range
$10^{11}<M_{\rm halo}/\msun<3\times10^{14}$ from a Sheth--Tormen halo mass
function \citep{1999MNRAS.308..119S}.\footnote{With $a=0.791$, $p=0.218$, and
normalisation $0.302$, the parameter values from \citet{2019MNRAS.485.5010B}}
Each $z=0$ host halo contains a central galaxy and may also host satellite
galaxies. The calculation follows the complete merger tree of each host,
including all progenitor branches.
The abundance of host haloes over this mass range corresponds to an effective comoving volume of $V_{\rm eff}\simeq 1.3\times10^6\,\mpc^3$, equivalent to a cubic box with side length $\simeq 109\,\mpc$.
Because the calibrated model is intended to generate \romanspace{}  mocks using the UNIT simulations \citep{2019MNRAS.487...48C}, we use a merger-tree mass resolution of $2\times10^{10}\,\msun$, similar to the minimum halo mass resolved in UNIT.
For the same reason, the merger trees are rebinned onto the UNIT simulation output times, following the approach of \citet{2012MNRAS.419.3590B}.

The use of extended Press--Schechter trees provides direct control over both
the number of simulated haloes and their mass distribution, allowing us to
obtain well-sampled predictions for the different calibration observables.
We verify in Section~\ref{sec:direct-map-validation} that the calibrated model gives
consistent predictions when applied instead to merger trees from the UNIT $N$-body
simulation.

\subsection{Calibrated Galacticus Parameters}
\label{sec:varied-parameters}

Our aim is to calibrate a model suitable for mock-catalogue generation, for which accurately reproducing a broad range of observed galaxy properties is more important than reserving observables as independent tests of the model. We therefore vary 20 \galacticus{} parameters and constrain them using several galaxy observables simultaneously.

This differs from the strategy often adopted for cosmological hydrodynamical
simulations, where a relatively small set of observables is used for
calibration so that other galaxy properties can serve as independent tests
of the model \citep[e.g.][]{2015MNRAS.446..521S, 2018MNRAS.475..624N, 2026MNRAS.548ag375S}. Our goal here is instead to obtain a model
that simultaneously reproduces the range of galaxy properties required for
the intended applications.

Our 20 free parameters control aspects of gas cooling, star formation, stellar feedback, black-hole seeding and growth, AGN feedback,
reincorporation of ejected gas, the classification of major and minor
mergers and the sizes of merger remnants, angular-momentum transfer during
disk instabilities, and the stellar metal yield. The parameters and their adopted ranges are listed in Table~\ref{tab:parameters}. The priors are intentionally broad and are generally centred on values used
in previous \galacticus{} configurations or suggested by earlier calibration
experiments, rather than representing precise prior constraints on the
underlying physical parameters.

\begin{table*}
\centering
\scriptsize
\caption{
Parameters varied in the emulator training campaign.
Here \(\mathcal{LN}_{[a,b]}(x_0,\sigma_{\ln})\) denotes a log-normal prior truncated to \([a,b]\), with median \(x_0\) and logarithmic width \(\sigma_{\ln}\), while \(\mathcal{U}_{[a,b]}\) denotes a uniform prior.
Units for dimensional parameters are shown in the parameter column, and priors are given for the numerical value in those units.
}
\label{tab:parameters}
\begin{tabular}{@{}p{0.21\textwidth}p{0.54\textwidth}p{0.21\textwidth}@{}}
\toprule
Parameter & Description & Prior \\
\midrule
\noalign{\vskip 1.8\baselineskip}
\multicolumn{3}{@{}l}{\normalfont Star formation and stellar populations} \\
\(\nu_{\rm SF,d}/\mathrm{Gyr}^{-1}\) & Disk star-formation frequency normalisation & \(\mathcal{LN}_{[0.1,10]}(1,2)\) \\
\(\epsilon_{\star,{\rm sph}}\) & Spheroid star-formation efficiency per dynamical time & \(\mathcal{LN}_{[10^{-4},1]}(0.01,4)\) \\
\(y_Z\) & Stellar metal yield & \(\mathcal{LN}_{[0.01,0.04]}(0.02,0.5)\) \\[0.8\baselineskip]
\multicolumn{3}{@{}l}{\normalfont Stellar feedback} \\
\(V_{\rm out,d}/\mathrm{km\,s^{-1}}\) & Disk stellar-feedback velocity scale & \(\mathcal{LN}_{[25,300]}(150,0.5)\) \\
\(\alpha_{\rm out,d}\) & Disk stellar-feedback velocity exponent & \(\mathcal{U}_{[0,4]}\) \\
\(V_{\rm out,sph}/\mathrm{km\,s^{-1}}\) & Spheroid stellar-feedback velocity scale & \(\mathcal{LN}_{[10,150]}(50,1)\) \\[0.8\baselineskip]
\multicolumn{3}{@{}l}{\normalfont Gas cooling and recycling} \\
\(f_{\rm cool}\) & Gas cooling-rate multiplier & \(\mathcal{LN}_{[0.05,1.2]}(0.5,1)\) \\
\(r_{\rm core}/R_{\rm vir}\) & Hot-gas core radius in units of virial radius & \(\mathcal{LN}_{[0.03,1]}(0.3,1)\) \\
\(\gamma_{\rm reinc}\) & Ejected-gas reincorporation normalisation & \(\mathcal{LN}_{[1,25]}(5,0.5)\) \\
\(\delta_{\rm reinc,1}\) & Redshift exponent in the \citet{2013MNRAS.431.3373H} reincorporation model & \(\mathcal{U}_{[-2,4]}\) \\
\(\delta_{\rm reinc,2}\) & Virial-velocity exponent in the \citet{2013MNRAS.431.3373H} reincorporation model & \(\mathcal{U}_{[-1,5]}\) \\[0.8\baselineskip]
\multicolumn{3}{@{}l}{\normalfont Mergers and disk instabilities} \\
\(f_{\rm major}\) & Mass-ratio threshold for major mergers & \(\mathcal{LN}_{[0.1,0.5]}(0.25,0.5)\) \\
\(f_{\rm orbit}\) & Orbital-energy factor for merger-remnant sizes & \(\mathcal{LN}_{[0.25,4]}(1,0.5)\) \\
\(f_{j,{\rm sph}}^{\rm bar}\) & Angular-momentum fraction retained by the spheroid in bar instabilities & \(\mathcal{LN}_{[0.05,0.95]}(0.2,0.5)\) \\[0.8\baselineskip]
\multicolumn{3}{@{}l}{\normalfont Supermassive black holes} \\
\(M_{\rm BH,seed}/\mathrm{M_\odot}\) & Black-hole seed mass & \(\mathcal{LN}_{[100,100\,000]}(3000,4)\) \\
\(\epsilon_{\rm BH,radio}\) & Radio-mode black-hole heating efficiency & \(\mathcal{LN}_{[0.03,1]}(0.3,0.5)\) \\
\(\epsilon_{\rm BH,wind}\) & Black-hole wind feedback efficiency & \(\mathcal{LN}_{[0.00024,0.024]}(0.0024,1)\) \\
\(\alpha_{\rm Bondi,sph}\) & Bondi accretion enhancement for spheroid gas & \(\mathcal{LN}_{[0.05,100]}(5,2)\) \\
\(\alpha_{\rm Bondi,hot}\) & Bondi accretion enhancement for hot-halo gas & \(\mathcal{LN}_{[0.05,100]}(6,2)\) \\
\(x_{\rm thin,max}\) & Upper transition value for $\dot{M}c^2/ L_{\rm Edd}$ for switched thin-disk/ADAF model & \(\mathcal{LN}_{[0.03,30]}(3,2)\) \\
\bottomrule
\end{tabular}
\end{table*}

\subsection{Emulator Training Set}
\label{sec:emulator-training-set}

The emulator training set consists of $N_{\rm train}=1024$ \galacticus{} model evaluations, with parameter values drawn from a scrambled Sobol sequence \citep{1967USSR...7...86S, 1997AnSta..25.1541O}.\footnote{The design was generated with \texttt{scipy.stats.qmc.Sobol}, using \texttt{scramble=True}, \texttt{seed=42}, and \texttt{random\_base2(m=10)}.}
Compared with random sampling, this provides a more uniform coverage of parameter space, reducing the likelihood of large unsampled regions and generally improving emulator accuracy for a fixed number of model evaluations.
We generate Sobol points in the unit hypercube and transform each coordinate through the inverse cumulative distribution function of the corresponding parameter prior.
As a result, the training density follows the prior density specified in Table~\ref{tab:parameters}, while retaining the space-filling properties of the Sobol design in prior-quantile space.
The choice of 1024 training evaluations represents a compromise between reducing emulator error across the 20-dimensional parameter space and keeping the computational cost of the \galacticus{} runs manageable.
Section~\ref{sec:smf-example} uses the low-redshift stellar mass function as a concrete example to illustrate how emulator accuracy changes with training-set size.

\subsection{Emission-Line And Dust Post-Processing}
\label{sec:dust-postprocessing}

A model that produces realistic emission-line luminosities is required for the \romanspace{} grism-survey application that motivates this calibration, because the \romanspace{} redshift sample will be selected primarily through rest-frame optical emission lines. We therefore include the \ha{} luminosity functions of \citet{2013MNRAS.428.1128S} among the calibration observables, requiring predictions for both intrinsic \ha{} emission and its attenuation by dust.

The \ha{} emission associated with star formation is produced mainly by
recombination in H\,{\sc ii} regions around young massive stars, and therefore
traces recent star formation \citep[e.g.][]{1998ARA&A..36..189K}. We calculate
this emission using the \galacticus{} implementation described by
\citet{2018MNRAS.474..177M}, which uses pre-computed \textsc{Cloudy}
\citep{2013RMxAA..49..137F} photoionization models together with the
star-formation and chemical-enrichment histories of each galaxy. For each
\galacticus{} run, we use the resulting intrinsic \ha{} luminosities stored in
the output catalogue, combining the contributions from star-forming regions in
the disk and spheroid with any contribution from the AGN.

These intrinsic luminosities cannot be compared directly with observed
luminosity functions, because \ha{} emission is attenuated by dust before
escaping the galaxy. We therefore apply a dust-attenuation model in
post-processing. As our baseline, we adopt the local relation between \ha{}
attenuation and stellar mass measured by \citet{2010MNRAS.409..421G}, which we
refer to as \gbten{}. This relation is based on Balmer-decrement measurements,
which estimate dust attenuation from ratios of hydrogen recombination lines.
Balmer-decrement measurements at higher redshift suggest that a mass-dependent
attenuation prescription remains a reasonable baseline, although some redshift
evolution is possible \citep[e.g.][]{2013ApJ...763..145D}. We therefore allow
departures from the \gbten{} relation in its normalisation and stellar-mass
dependence, together with redshift evolution and mass-dependent redshift
evolution.

For each \galacticus{} training run, emission-line luminosity functions are computed after applying this dust model. The five dust parameters, which describe departures from the \gbten{} relation and the scatter about it, are sampled jointly with the 20 \galacticus{} parameters in the final inference. The mathematical form of the attenuation model, its parameter priors, and the construction of the attenuated \ha{} luminosity functions are described in Appendix~\ref{app:dust-model-details}.

\section{Emulating Calibration Observables}
\label{sec:emulator}

\subsection{Set Of Calibration Observables}
\label{sec:calibration-observable-set}

Each \galacticus{} evaluation produces a catalogue of galaxies at a set of
output times. From these galaxies we compute binned observable vectors, such as
stellar mass functions, luminosity functions, size--mass relations, and black hole scaling
relations. The emulator is trained to reproduce these vectors as a function of
the varied \galacticus{} parameters.

The final calibration uses the observable set listed in
Table~\ref{tab:calibration-observable-summary}. For number-density observables
we denote the binned differential number density by $\Phi$, with the argument
indicating the quantity being binned, for example $\Phi(M_\star)$ for a stellar
mass function and $\Phi(L_{\mathrm{H}\alpha})$ for an \ha{} luminosity function.
The emulated-quantity column of the table gives the transformed quantity
passed to the emulator and used in the likelihood. %This is also the space in which the emulator-validation RMSE values in Table~\ref{tab:emulator-validation-summary} are measured. For example, an RMSE of 0.1 for a luminosity or mass function means 0.1 dex in $\log_{10}\Phi$, not a linear number-density error.

For each calibration observable, we also estimate the finite-sampling
uncertainty associated with constructing the binned prediction from a finite
\galacticus{} galaxy catalogue. For number-density observables, this corresponds
to the counting uncertainty associated with the galaxies contributing to each
bin. For binned relations, such as the size--mass,
black-hole--velocity-dispersion, and mass--metallicity relations, the emulated
quantity is the mean transformed galaxy property in each bin, and the
finite-sampling uncertainty corresponds to the uncertainty on this mean.
These per-bin uncertainties are included during Gaussian-process training as a
diagonal training-noise term, so that noisier training outputs are treated as
less precise estimates of the underlying model prediction.

\begin{table*}
\centering
\caption{
Calibration observables used in this work.
The final joint fit uses all listed observables.
The emulated quantity defines the space used for emulator training and posterior
inference, and is the space in which validation RMSE values are computed.
\(N_{\rm bin}\) is the number of binned data-vector elements entering the
emulator and likelihood for the corresponding row.
\(N_{\rm PC}\) is the number of principal components required to explain
99 per cent of the variance in the standardised training set.
}
\label{tab:calibration-observable-summary}
\begin{tabular}{@{}p{0.28\textwidth}p{0.18\textwidth}p{0.17\textwidth}p{0.12\textwidth}rr@{}}
\toprule
Calibration observable
  & Reference data
  & Emulated quantity
  & Redshift
  & \(N_{\rm bin}\)
  & \(N_{\rm PC}\) \\
\midrule

\multirow{2}{0.28\textwidth}{Stellar mass function}
  & \multirow{2}{0.17\textwidth}{\citet{2014ApJ...783...85T}}
  & \multirow{2}{0.17\textwidth}{\(\log_{10}\Phi(M_\star)\)}
  & \(0.20<z<0.50\) & 14 & 6 \\
  & & & \(1.00<z<1.25\) & 11 & 4 \\
\addlinespace[0.5em]

Star-formation-rate function
  & \citet{2011MNRAS.416.2640R}
  & \(\log_{10}\Phi(\dot{M}_\star)\)
  & \(0.013<z<0.1\) & 25 & 7 \\
\addlinespace[0.5em]

Size--mass relation (star-forming)
  & \multirow{2}{0.17\textwidth}{\citet{2014ApJ...788...28V}}
  & \(\log_{10}R_{\rm eff}(M_\star)\)
  & \(0<z<0.5\) & 4 & 3 \\

Size--mass relation (quiescent)
  &
  & \(\log_{10}R_{\rm eff}(M_\star)\)
  & \(0<z<0.5\) & 5 & 4 \\
\addlinespace[0.5em]

Black-hole--velocity-dispersion relation
  & \citet{2013ApJ...764..184M}
  & \(\log_{10}M_{\rm BH}(\sigma_\star)\)
  & \(z\approx0\) & 5 & 4 \\
\addlinespace[0.5em]

Gas-phase mass--metallicity relation
  & \citet{2019ApJ...877....6B}
  & \(12+\log_{10}({\rm O/H})\)
  & \(z\approx0\) & 49 & 6 \\
\addlinespace[0.5em]

\multirow{4}{0.28\textwidth}{\ha{} luminosity function}
  & \multirow{4}{0.17\textwidth}{\citet{2013MNRAS.428.1128S}}
  & \multirow{4}{0.17\textwidth}{\(\log_{10}\Phi(L_{\mathrm{H}\alpha})\)}
  & \(z=0.40\) & 18 & 6 \\
  & & & \(z=0.84\) & 9 & 5 \\
  & & & \(z=1.47\) & 13 & 7 \\
  & & & \(z=2.23\) & 15 & 8 \\
\bottomrule
\end{tabular}
\end{table*}

\subsubsection{Observable-Space Transformations}
\label{sec:observable-space-transformations}

For observables that span a large dynamic range across \galacticus{} parameter
space, we emulate logarithmic quantities.  For number-density observables this
means emulating $\log_{10}\Phi$ rather than $\Phi$ itself.  At fixed stellar
mass, star formation rate, or emission-line luminosity, changes in the
\galacticus{} parameters can produce order-of-magnitude changes in the
predicted number density.  The logarithmic transformation therefore gives a smoother emulator target across parameter space and makes emulator errors more closely related to fractional differences in $\Phi$. It also matches the space in which luminosity and mass functions are usually compared.

We use the same approach for other positive quantities with broad dynamic
range.  The size--mass relations are emulated as $\log_{10}R_{\rm eff}$ and
the black-hole scaling relation as $\log_{10}M_{\rm BH}$.  The gas-phase
mass--metallicity relation is already conventionally expressed as
$12+\log_{10}({\rm O/H})$, and is emulated in that form.

The transformations require special treatment when a training-set prediction
is zero or undefined. For example, a model may produce no galaxies in a
stellar-mass-function bin, so that $\log_{10}\Phi$ is undefined, or no
quiescent galaxies in a high-stellar-mass bin, so that the binned mean size is
undefined. For number densities, we impose a small positive floor before taking the logarithm. For undefined binned relations, we replace the missing value by the median valid training-set value in that bin and assign it a large training uncertainty, so that the assigned value carries very little weight when fitting the emulator. The detailed floor and missing-value procedure is described in Appendix~\ref{app:missing-values}.

\subsubsection{Principal-Component Compression}
\label{sec:pca-compression}

A simple approach to emulating each observable would be to build a separate emulator for every bin: for example, one emulator for the number density of galaxies at $M_\star\simeq10^{10}\msun$, another at $M_\star\simeq2\times10^{10}\msun$, and so on.
This is straightforward, but it treats neighbouring bins as independent targets even though changes in \galacticus{} parameters usually move the predictions coherently across an observable.
For example, a change in feedback strength will typically shift the amplitude, slope, or knee of a stellar mass function, rather than perturbing each stellar-mass bin independently.

We therefore use principal-component analysis (PCA) to compress each observable vector before emulation.
This reduces the number of emulators that must be trained and evaluated, and can improve predictive accuracy by exploiting correlations in how the different bins of an observable vary across parameter space.
Principal-component compression is commonly used when emulating expensive simulation outputs with many correlated bins, including recent applications to cosmological and galaxy-formation inference \citep[e.g.][]{2026arXiv260107306R}.
For a given observable, let $\boldsymbol{\theta}$ denote the relevant model parameters: the varied \galacticus{} parameters and, where applicable, the dust-model parameters. We write
\begin{equation}
  \mathbf{y}(\boldsymbol{\theta})
  =
  \left(y_1(\boldsymbol{\theta}), \ldots, y_{N_{\rm bin}}(\boldsymbol{\theta})\right)
\end{equation}
for the vector of predictions in the $N_{\rm bin}$ bins used for that observable.
For the stellar mass function, for example, $y_b(\boldsymbol{\theta})=\log_{10}\Phi_b(\boldsymbol{\theta})$, where $\Phi_b$ is the predicted number density in the $b$th stellar-mass bin.
For each \galacticus{} run in the training set we have one such vector, so the training data for this observable form an $N_{\rm train}\times N_{\rm bin}$ matrix; we use $n$ to index the individual \galacticus{} runs in the training set.

We first standardise each bin using its mean, $\mu_b$, and
standard deviation, $s_b$, across the training-set runs. For training-set run $n$, the standardised value in
bin $b$ is
\begin{equation}
  z_{nb} =
  \frac{y_b(\boldsymbol{\theta}_n)-\mu_b}{s_b}.
\end{equation}
We then apply PCA to the resulting $N_{\rm train}\times N_{\rm bin}$ matrix
with elements $z_{nb}$.
PCA defines an orthonormal set
of directions, $\mathbf{e}_k$, in observable-bin space, ordered by the variance
of the training set projected along each direction \citep[e.g.][]{2016RSPTA.37450202J}.
For training model $n$, its coefficient along component $k$ is
\begin{equation}
  a_{nk} = \mathbf{e}_k\cdot\mathbf{z}_n
         = \sum_b e_{kb} z_{nb}.
\end{equation}
We retain the smallest number of components, $K$, required to explain 99 per
cent of the variance in the standardised training set.
The quantities $a_{nk}$ are then the training targets for the emulator associated with component $k$; after training, the emulator predicts the corresponding coefficient, $a_k(\boldsymbol{\theta})$, at new parameter values.
The predicted observable vector in standardised observable space is reconstructed as
\begin{equation}
  \mathbf{z}(\boldsymbol{\theta}) \simeq
  \sum_{k=1}^{K} a_k(\boldsymbol{\theta}) \mathbf{e}_k.
\end{equation}
Finally, we undo the bin-wise standardisation to recover the emulator target in each bin,
\begin{equation}
  y_b(\boldsymbol{\theta}) \simeq
  \mu_b + s_b z_b(\boldsymbol{\theta}).
\end{equation}

For comparison and diagnostic plots, we also train bin-by-bin emulators.
The main calibration results use PCA emulators with a 99 per cent variance threshold; the cross-validation tests below show why this is a reasonable choice. The number of retained PCA components for each calibration observable is listed in Table~\ref{tab:calibration-observable-summary}.

\subsection{Gaussian-Process Emulators}
\label{sec:gp-regression}

Each retained PCA coefficient is modelled with a Gaussian process (GP) as a function of the input parameters.
The same basic GP choices are used for the emulators compared below, so that the validation tests mainly probe the effects of training-set size and PCA compression.
For the 20 \galacticus{} parameters, the GP inputs are not the physical parameter values themselves, but the corresponding prior quantiles.
That is, for \galacticus{} parameter $\theta_i$ with prior cumulative distribution function $P_i(<\theta_i)$, we train the emulator using
\begin{equation}
  u_i = P_i(<\theta_i),
\end{equation}
so that each input parameter is mapped onto the unit interval.
Because the Sobol design was constructed in this space, the training points
provide approximately uniform coverage in prior-quantile coordinates.
It also avoids asking the GP to learn across raw parameters with very different units, ranges, and prior shapes; an interval of fixed width in $u_i$ contains the same prior probability for each parameter.
The \ha{} luminosity-function emulator has five additional inputs: the dust post-processing parameters described in Section~\ref{sec:dust-postprocessing}.

After the PCA transformation, each retained coefficient is standardised once more before fitting its GP, by subtracting its training-set mean and dividing by its training-set standard deviation.
This is separate from the bin-wise standardisation applied before PCA, and is undone before reconstructing the observable vector.
Let $\mathbf{u}$ denote the complete input vector supplied to a given GP after the input transformations just described.
The kernel specifies the covariance between the values of the function at two points in this input space.
We use a Mat\'ern kernel with $\nu=2.5$ \citep[e.g.][]{2006gpml.book.....R},
multiplied by a constant amplitude, and allow a separate length scale for each
input coordinate.
For two input vectors, $\mathbf{u}$ and $\mathbf{u}'$, the kernel is
\begin{equation}
  k(\mathbf{u},\mathbf{u}')
  =
  A
  \left(1+\sqrt{5}r+\frac{5r^2}{3}\right)
  \exp\left(-\sqrt{5}r\right),
\end{equation}
where
\begin{equation}
  r^2 = \sum_i \left(\frac{u_i-u_i'}{\ell_i}\right)^2.
\end{equation}
Here $A$ is the covariance amplitude and $\ell_i$ is the length scale associated with coordinate $u_i$ of the transformed input space.
Small values of $\ell_i$ allow the emulator prediction to vary rapidly as that input coordinate is changed, while large values correspond to smoother variation.
For each PCA coefficient, the kernel hyperparameters are therefore the overall covariance amplitude and one length scale for each emulator input.
We determine the kernel hyperparameters by maximising the GP marginal
likelihood \citep{2006gpml.book.....R} using a gradient-based optimiser, starting from multiple initial points to reduce sensitivity to local maxima.

\subsubsection{Emulator Uncertainties}

The observables measured from each \galacticus{} training run have
finite-sampling uncertainties, arising from the finite number of galaxies used
to estimate each binned quantity. We account for these uncertainties during GP
training by adding their variances to the diagonal of the covariance matrix
obtained by evaluating the kernel at the training points
\citep[e.g.][]{2006gpml.book.....R}. The training uncertainty can therefore vary between individual training models and observable bins.
For PCA emulators, the
per-bin uncertainties are propagated through the standardisation and PCA
transformations to obtain a training uncertainty for each retained PCA
coefficient. We neglect correlations between observable bins in this
propagation. Further details are given in
Appendix~\ref{app:covariance-treatment}.

At a new point in parameter space, the GP provides both a mean prediction and
a predictive variance that quantifies the emulator's uncertainty in that
prediction. We propagate this predictive variance into the calibration
likelihood by combining it with the observational uncertainty (see
Section~\ref{sec:likelihood}). This requires the emulator uncertainties to be
reasonably calibrated, making the held-out tests in
Section~\ref{sec:emulator-validation} an important part of assessing the
reliability of the inference.

\subsubsection{Held-Out Emulator Validation}
\label{sec:emulator-validation}

We validate the emulator using \emph{five-fold cross-validation}.
The \galacticus{} training runs are divided into five equally sized subsets.
For each fold, the emulator is trained on four subsets and evaluated on the remaining subset, so that every \galacticus{} evaluation is predicted once by an emulator that was not trained using that model.
Unless otherwise stated, all validation results in this section are based on the combined predictions from these five folds.

These held-out predictions test whether the emulator can reproduce direct \galacticus{} calculations at new parameter values, and allow us to assess how choices such as training-set size and PCA compression affect its predictive accuracy.
A summary of the held-out performance for all calibration-observable emulators is given in Appendix~\ref{app:emulator-validation-summary}, where we also account explicitly for the finite-sampling uncertainty of the held-out \galacticus{} calculations.
Section~\ref{sec:smf-example} illustrates the validation procedure for the low-redshift stellar mass function.

To quantify this comparison, we consider individual bins of the observable vector.
For a given bin, we compare the emulator prediction, $m_i$, with the direct \galacticus{} value, $g_i$, for each validation model $i$.
The root-mean-square error,
\begin{equation}
  {\rm RMSE} = \left[\frac{1}{N_{\rm val}}\sum_i (m_i-g_i)^2\right]^{1/2},
\end{equation}
measures the typical difference between the emulator and direct \galacticus{} predictions.
We also use the coefficient of determination,
\begin{equation}
  R^2 = 1 - \frac{\sum_i (m_i-g_i)^2}{\sum_i (g_i-\bar{g})^2},
\end{equation}
where $\bar{g}$ is the mean direct \galacticus{} value over all validation models.
This compares the emulator error with the variation of the direct \galacticus{} predictions in that bin: $R^2=1$ corresponds to perfect predictions, while $R^2=0$ corresponds to doing no better than predicting the mean \galacticus{} value for all parameter choices.
For the stellar mass functions shown below, both $m_i$ and $g_i$ are values of $\log_{10}\Phi$ in a fixed stellar-mass bin, so the RMSE is measured in dex.

\subsection{Low-Redshift Stellar Mass Function Example}
\label{sec:smf-example}

To give some intuition for the validation procedure and metrics discussed above,
we now consider one concrete example: the low-redshift stellar mass function.
We show the emulator predictions alongside the corresponding held-out
\galacticus{} results, and then examine the validation metrics and their
dependence on the construction and size of the training set.

\begin{figure*}
\centering
\includegraphics[width=0.85\textwidth]{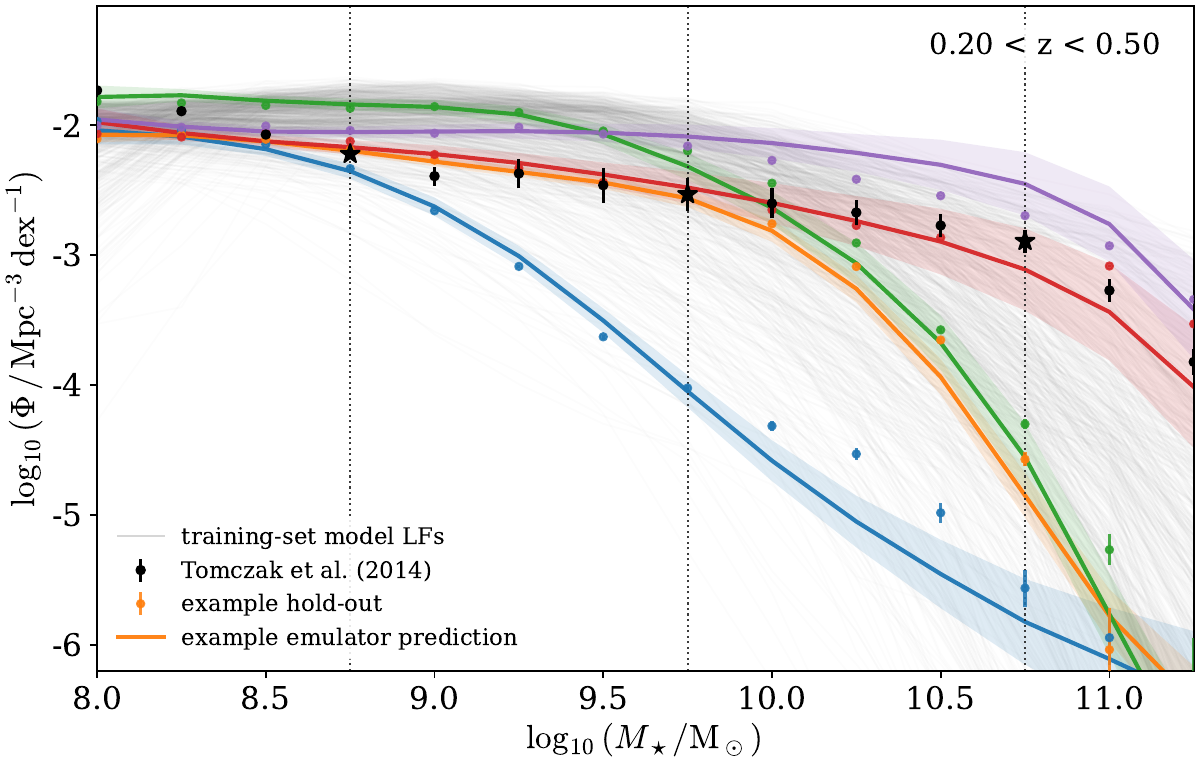}
\caption{
Example of held-out validation for the low-$z$ stellar mass function.
Thin grey curves show the range of \galacticus{} predictions across the training set.
Black points show the observed stellar mass function from \citet{2014ApJ...783...85T}.
Coloured points show direct \galacticus{} predictions for example held-out parameter choices.
The corresponding coloured curves show emulator predictions for the same parameter choices, where in each case the emulator was trained without using the corresponding \galacticus{} evaluation.
The shaded regions show the emulator uncertainty.
Five held-out models spanning the range of predicted stellar mass functions are selected systematically for visual clarity; the validation metrics use the full held-out sample.
Vertical dotted lines mark the stellar-mass bins used for the bin-level validation metrics in Figures~\ref{fig:example_predicted_vs_heldout} and \ref{fig:training_size}.
}
\label{fig:SMF-holdout-demo}
\end{figure*}

Figure~\ref{fig:SMF-holdout-demo} illustrates the held-out validation procedure
for the low-redshift stellar mass function. The emulator reproduces the broad
variation in stellar mass functions across the held-out models, with the
largest discrepancies occurring towards the high-mass end and for models with
more extreme mass functions.

\begin{figure*}
\centering
\includegraphics[width=0.85\textwidth]{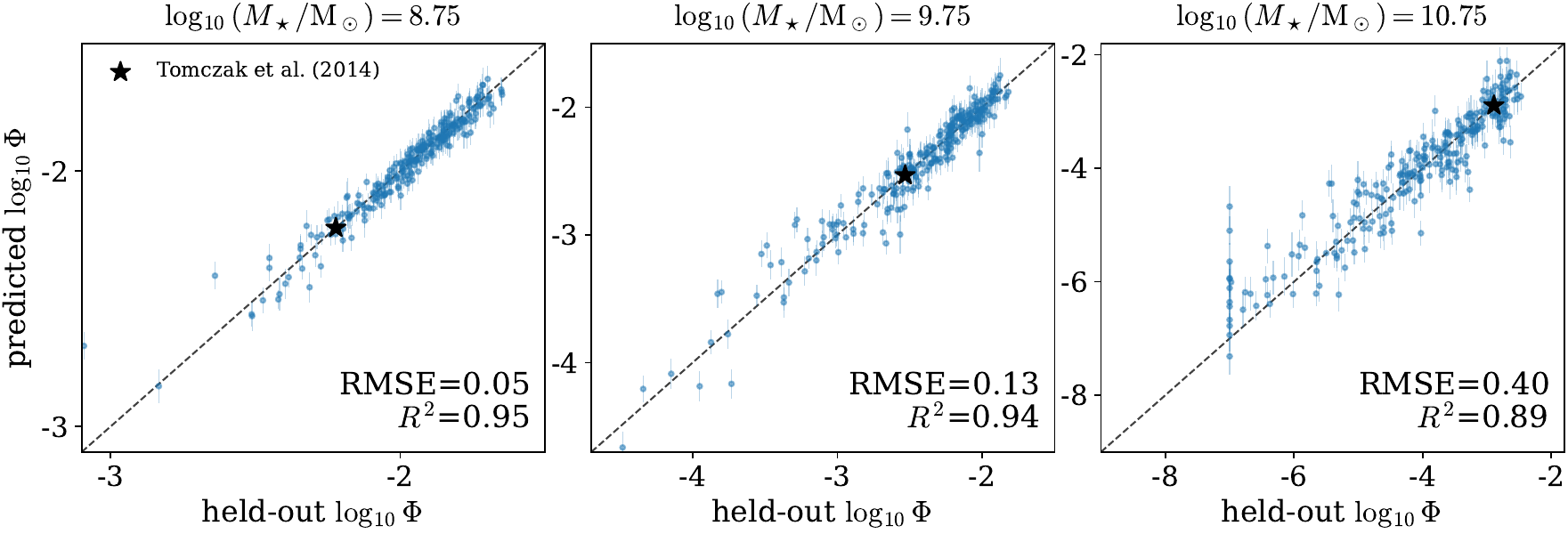}
\caption{
Predicted versus held-out values of the low-$z$ stellar mass function at the three stellar-mass bins highlighted in Figure~\ref{fig:SMF-holdout-demo}.
The $x$-axis values are the direct \galacticus{} predictions, while the $y$-axis shows the emulator predictions (with their uncertainty) at the same parameter values. The dashed lines show equality.
The RMSE and $R^2$ values quoted in each panel summarise the accuracy of these held-out predictions, using the definitions given in Section~\ref{sec:emulator-validation}.
For clarity, the figure shows a random 25 per cent of the held-out comparisons.
}
\label{fig:example_predicted_vs_heldout}
\end{figure*}

Figure~\ref{fig:example_predicted_vs_heldout} shows this bin-level comparison for three representative stellar-mass bins from Figure~\ref{fig:SMF-holdout-demo}.
The plotted points should lie close to the one-to-one line if the emulator is accurate.
The RMSE values measure the absolute scatter about this line, while the $R^2$ values measure how small that scatter is relative to the variation in the direct \galacticus{} predictions.

\begin{figure*}
\centering
\includegraphics[width=0.85\textwidth]{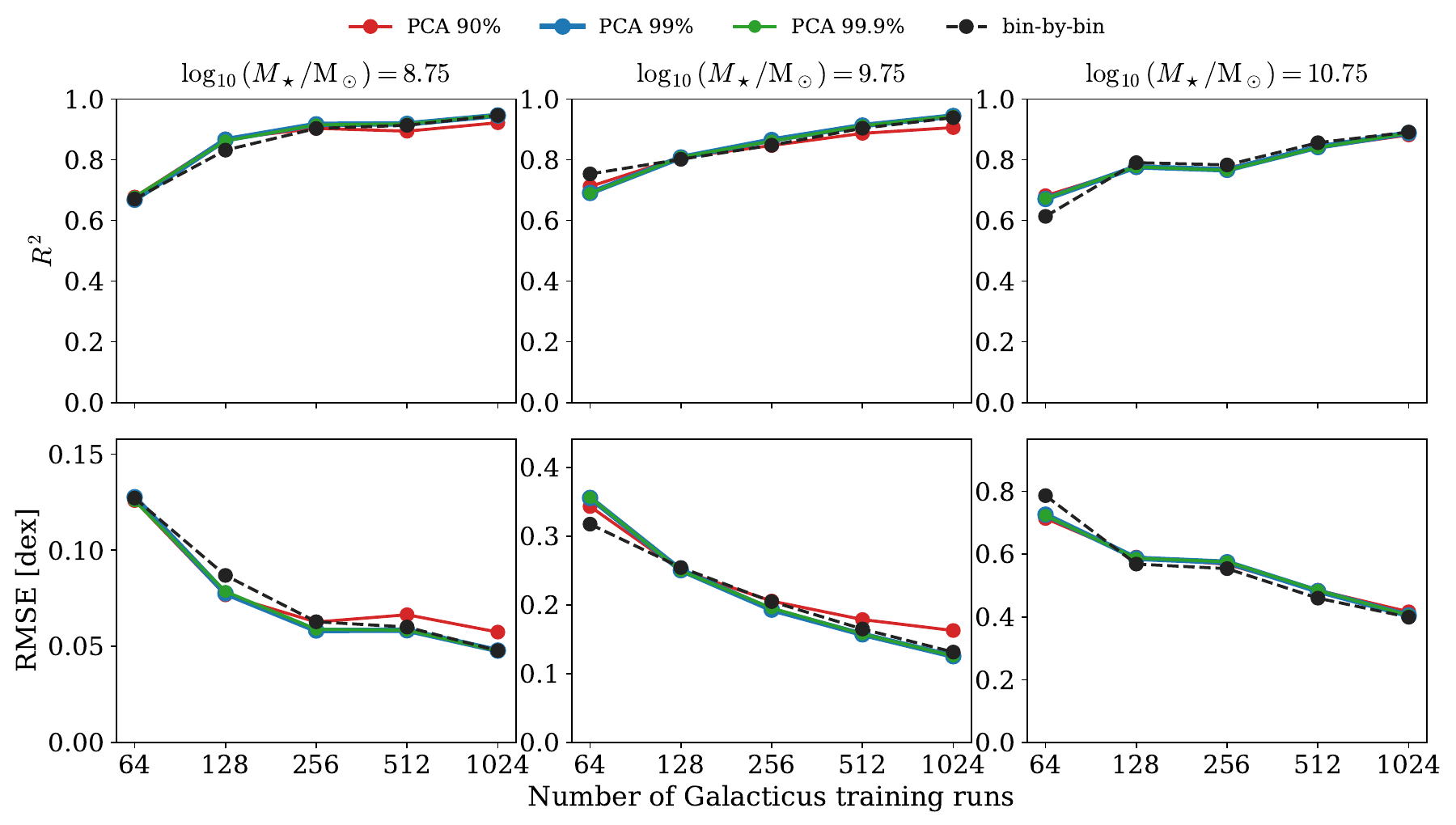}
\caption{
Held-out emulator accuracy for the $z\simeq0$ stellar mass function as a function of the number of \galacticus{} training runs.
The top row shows the coefficient of determination, $R^2$, and the bottom row shows the root-mean-square error in $\log_{10}\Phi$.
The columns correspond to the representative stellar-mass bins highlighted in Figure~\ref{fig:SMF-holdout-demo}.
Different curves compare PCA-GP emulators with different PCA variance thresholds to a bin-by-bin GP emulator.
The improvement from smaller training subsets to the full 1024-run training set motivates the larger final campaign used for the paper calibration, while the similarity of the PCA-GP variants supports the 99 per cent PCA variance threshold used in the fiducial emulator.
}
\label{fig:training_size}
\end{figure*}

Figure~\ref{fig:training_size} shows how these diagnostics depend on training-set size and emulator construction for several representative stellar-mass bins.
We compare emulators trained on the first 64, 128, 256, 512, and 1024 points of the Sobol sequence.
Because the first $2^m$ points of a Sobol sequence are themselves designed to provide good coverage of parameter space, these subsets provide a natural sequence of training sets for assessing the effect of increasing the training-set size.
Emulator accuracy improves substantially as the training set grows from 64 to 1024 models, although the improvement becomes smaller for the larger training sets, suggesting diminishing returns from further increasing the training-set size.
We therefore adopt the 1024-run training set as a practical compromise between emulator accuracy and computational cost.

Figure~\ref{fig:training_size} also compares different levels of PCA
compression. Retaining 90 per cent of the PCA variance leads to somewhat
poorer held-out accuracy, while the 99 and 99.9 per cent thresholds give very
similar results. Their performance is also comparable to, and in some cases
slightly better than, the bin-by-bin emulators. We therefore adopt the 99 per
cent threshold for the fiducial emulator, since retaining additional PCA
components produces no clear improvement in predictive accuracy.

\section{Posterior Inference Of Model Parameters}
\label{sec:posterior-inference}

\subsection{MCMC Fits Used In This Work}
\label{sec:mcmc-fits}

We use two kinds of MCMC fit.
First, we run a small set of diagnostic fits to subsets of the calibration data to illustrate how different observables constrain different combinations of \galacticus{} parameters.
The example shown in Section~\ref{sec:degeneracy-breaking} uses stellar mass functions, size--mass relations, and the star-formation-rate function, both separately and jointly.
Second, we run the final joint calibration using the full set of calibration observables summarised in Table~\ref{tab:calibration-observable-summary}.
The final emulators are trained on the 1024-run training set described in Section~\ref{sec:emulator-training-set}. The final joint fit samples the 20 \galacticus{} parameters in Table~\ref{tab:parameters} together with the five dust parameters described in Appendix~\ref{app:dust-model-details}.

We infer the model parameters from the posterior distribution
\begin{equation}
  p(\boldsymbol{\theta}\mid\mathbf{d})
  \propto
  \mathcal{L}(\mathbf{d}\mid\boldsymbol{\theta})
  p(\boldsymbol{\theta}),
\end{equation}
where $p(\boldsymbol{\theta})$ is the product of the parameter priors specified
in Table~\ref{tab:parameters} and, for the joint calibration, the dust-parameter
priors given in Appendix~\ref{app:dust-model-details}.

\subsection{The Likelihood}
\label{sec:likelihood}

The MCMC likelihood is evaluated in the same transformed observable space in
which the emulators are trained.  For stellar mass functions, the star
formation rate function, and the \ha{} luminosity functions, this is
$\log_{10}$ number density.  For the size--mass relations and black-hole
relation, the fitted quantities are already logarithmic quantities,
$\log_{10} R_{\rm eff}$ and $\log_{10} M_{\rm BH}$, respectively.  For the
mass--metallicity relation, the fitted quantity is $12+\log_{10}({\rm O/H})$.

For each observable bin $i$, we assume a Gaussian likelihood in this
transformed space,
\begin{equation}
  \ln \mathcal{L}
  =
  -\frac{1}{2}
  \sum_i
  \left[
    \frac{\left(d_i - m_i(\boldsymbol{\theta})\right)^2}
         {s_i^2(\boldsymbol{\theta})}
    +
    \ln\left(2\pi s_i^2(\boldsymbol{\theta})\right)
  \right],
\end{equation}
where $d_i$ is the observed value, $m_i(\boldsymbol{\theta})$ is the emulator
mean prediction, and the total variance is
\begin{equation}
  s_i^2(\boldsymbol{\theta})
  =
  \sigma_{{\rm data},i}^2
  +
  \sigma_{{\rm emu},i}^2(\boldsymbol{\theta}).
\end{equation}
Here $\sigma_{{\rm data},i}$ is the observational uncertainty and
$\sigma_{{\rm emu},i}(\boldsymbol{\theta})$ is the GP predictive uncertainty.
The observational and emulator uncertainties are therefore added in quadrature
in the transformed observable space.

We treat the uncertainties in different observable bins as independent, using
only the diagonal observational and emulator variances in the likelihood.
We therefore neglect correlations both between bins of a given observable and
between different calibration observables.

For number-density datasets whose uncertainties are provided in linear space,
we propagate the diagonal uncertainties to logarithmic number density using
\begin{equation}
  \sigma_{\log_{10}\Phi}
  \simeq
  \frac{\sigma_\Phi}{\Phi\ln 10}.
\end{equation}
For the \citet{2013MNRAS.428.1128S} \ha{} luminosity functions, the published
uncertainties are already given in $\log_{10}\Phi$, so no conversion from linear
number density is required. We use these tabulated values directly as symmetric
Gaussian uncertainties in the transformed observable space. This approximation is most appropriate when the fractional
uncertainties are moderate. For sparsely populated number-density bins, the underlying counting uncertainty is intrinsically asymmetric and is only approximately represented in this way.

\subsection{MCMC Sampling And Convergence}
\label{sec:mcmc-sampling}

We sample the posterior distribution using Markov chain Monte Carlo (MCMC), implemented with the affine-invariant ensemble sampler \emcee{} \citep{ForemanMackey:2013io}.
Each MCMC uses 128 walkers, initialised in a small region around a point expected to lie close to the maximum of the posterior density, which was determined from a preliminary MCMC.
During burn-in, the walkers disperse from this region and explore the posterior distribution. Initialising the walkers in this way helps reduce the likelihood of walkers becoming trapped around low-probability local maxima.

For the production runs, we use 250,000 steps per walker, discarding the initial 25,000 steps as burn-in.
We assess convergence using the integrated autocorrelation time, $\tau$ \citep{Sokal1997}, estimated separately for each sampled parameter.
For the diagnostic fits used to illustrate parameter degeneracies (see Fig.~\ref{fig:combining_constraints}), the post-burn-in chains extend for at least 25 times the largest measured autocorrelation time.
For the final joint calibration, the chain length is 73 times the largest measured autocorrelation time.
We therefore consider the chains sufficiently long for the posterior inferences presented here.

The emulator predictions are used in place of direct \galacticus{} evaluations.
This dramatically speeds up inference, because calling the emulator takes a few milliseconds, while a typical \galacticus{} evaluation takes approximately 15 hours using 16 CPU cores.\footnote{Generating the 1024-run emulator training set required approximately 250,000 CPU hours in total.}

\section{Calibration Results}
\label{sec:calibration-results}

\subsection{Breaking Parameter Degeneracies With Multiple Observables}
\label{sec:degeneracy-breaking}

\begin{figure*}
\centering
\includegraphics[width=\textwidth,trim={0 1cm 0 0},clip]{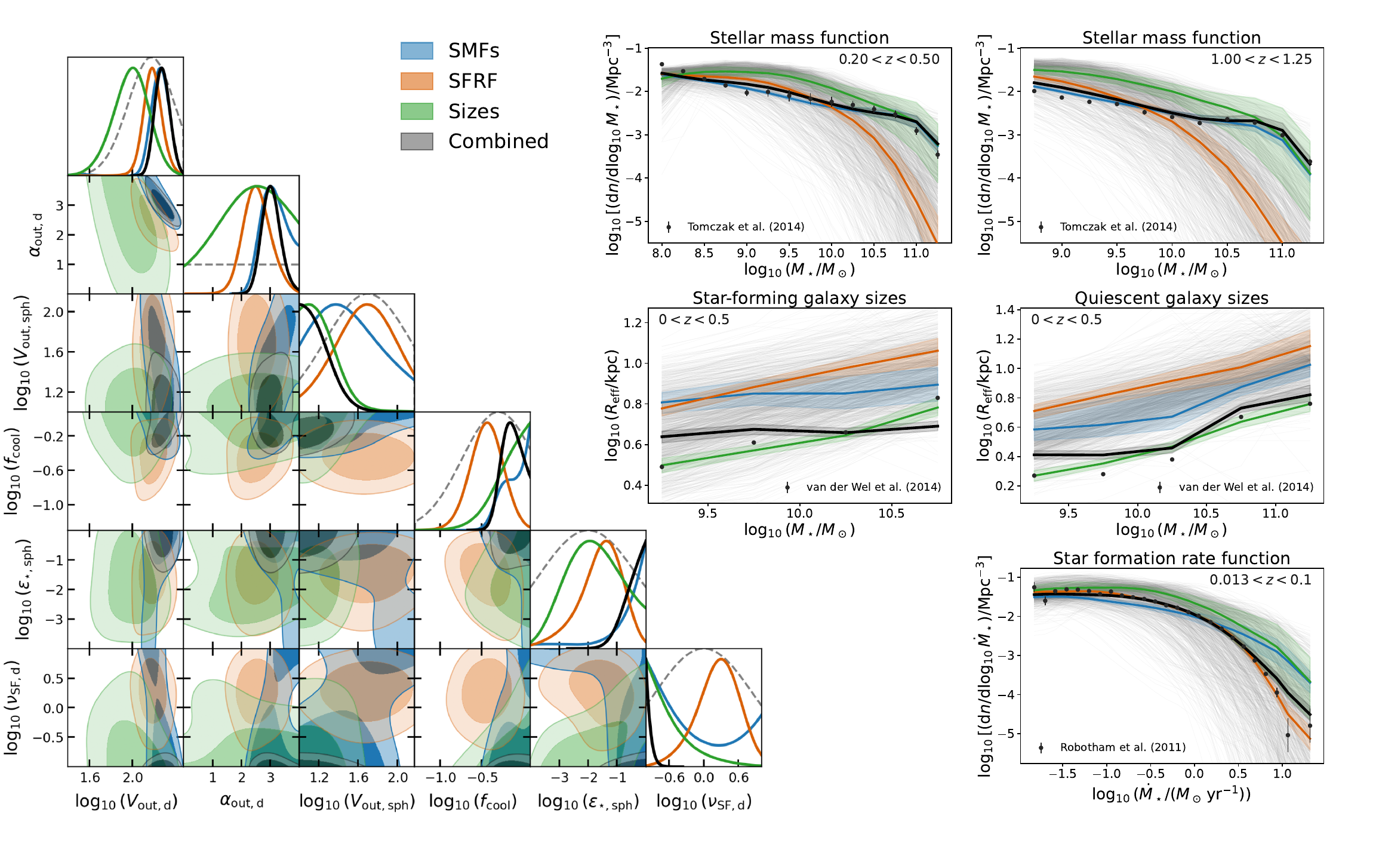}
\caption{
An example of combining multiple observables to break degeneracies in the
emulator calibration.
Left: marginalised posterior constraints on a subset of six model parameters
from fits to the stellar mass functions alone, the star-formation-rate
function alone, galaxy sizes alone, and their combination.
Diagonal panels show the one-dimensional marginalised posteriors, with grey
dashed curves showing the adopted priors.
Off-diagonal panels show the corresponding two-dimensional marginalised
posterior distributions, with 68 and 95 per cent credible regions.
Right: solid coloured curves and shaded regions show the median and central
68 per cent intervals of the emulator-predicted mean relations obtained by
drawing parameters from each posterior, compared with the observational data.
These intervals reflect posterior uncertainty in the calibrated model
parameters and do not include observational noise or emulator uncertainty.
Faint grey curves show the predictions from the 1024 training runs,
illustrating the range spanned by the prior over model parameters.
The combined calibration achieves reasonable agreement with all fitted observables
while reducing the parameter degeneracies present in the single-observable
fits.
}
\label{fig:combining_constraints}
\end{figure*}

To build intuition for how different observables constrain the model, we consider a set of diagnostic calibrations using the stellar mass functions, the star-formation-rate function, and stellar mass--size relations. These provide a controlled example of how combining observables can reduce parameter degeneracies and reveal tensions between different aspects of the galaxy population.

In each diagnostic fit, all 20 \galacticus{} parameters listed in Table~\ref{tab:parameters} are varied. For visual clarity, Fig.~\ref{fig:combining_constraints} shows only a subset of six parameters, chosen because they control several of the main processes regulating the stellar component. These are the normalisation and circular-velocity scaling of stellar feedback in disks, $V_{\rm out,d}$ and $\alpha_{\rm out,d}$; the corresponding feedback velocity scale in spheroids, $V_{\rm out,sph}$; the overall gas cooling rate, $f_{\rm cool}$; and the parameters setting the rate at which gas is converted into stars in spheroids and disks, $\epsilon_{\star,{\rm sph}}$ and $\nu_{\rm SF,d}$. We focus on these six parameters for illustration, although the remaining parameters, including those controlling the recycling
of ejected gas, black-hole growth, and feedback from accreting black holes, also
affect the stellar mass and star-formation histories.

The left-hand part of Fig.~\ref{fig:combining_constraints} shows that
calibrations to individual observables leave substantial parameter
degeneracies. The stellar mass functions, for example, can be reproduced
by different combinations of gas cooling, star-formation efficiency, and
feedback strength. They also permit a trade-off between disk and spheroid
feedback, visible in the $V_{\rm out,sph}$--$V_{\rm out,d}$ plane: stronger
feedback in one component (larger $V_{\rm out}$) can be partially compensated
by weaker feedback in the other. The star-formation-rate function and
size--mass relations select different regions and degeneracy directions in
the same parameter space. Combining the observables therefore substantially
tightens the constraints on several of the parameters shown.

The size--mass relations provide a particularly clear example of both the
complementarity and the tension between the observables. Over much of the
prior volume, \galacticus{} produces galaxies that are too large compared
with the observed relations. Matching the observed sizes favours weaker
stellar feedback. This enables more effective star formation in lower-mass
haloes, shifting galaxies of a given stellar mass towards less massive haloes
\citep{2009MNRAS.397.1254G,2016MNRAS.462.3854L}. Since galaxy sizes in
semi-analytic models are linked to the sizes and angular momenta of their
host dark matter haloes \citep[e.g.][]{2000MNRAS.319..168C}, this tends to
produce smaller galaxies at fixed stellar mass.

The same shift in the relation between stellar mass and halo mass has
an immediate consequence for the stellar mass functions: galaxies form more
stellar mass at fixed halo mass, shifting the predicted stellar mass functions
towards higher abundances. In Fig.~\ref{fig:combining_constraints}, the weaker
feedback favoured by the size--mass relations improves the predicted galaxy
sizes, but produces stellar mass functions that lie above the observations.
The combined fit therefore represents a compromise, selecting a region of
parameter space that gives good simultaneous agreement with the stellar mass
functions, star-formation-rate function, and galaxy sizes.

This example illustrates the benefit of calibrating to multiple observables.
Different observables respond to different combinations of physical
parameters, so fitting them together can constrain directions in parameter
space that remain degenerate when any one observable is considered in
isolation. We now apply this approach to the full set of observables used in
our final calibration.

\subsection{Final Joint Calibration}
\label{sec:final-joint-calibration}

\begin{figure*}
\centering
\includegraphics[width=\textwidth]{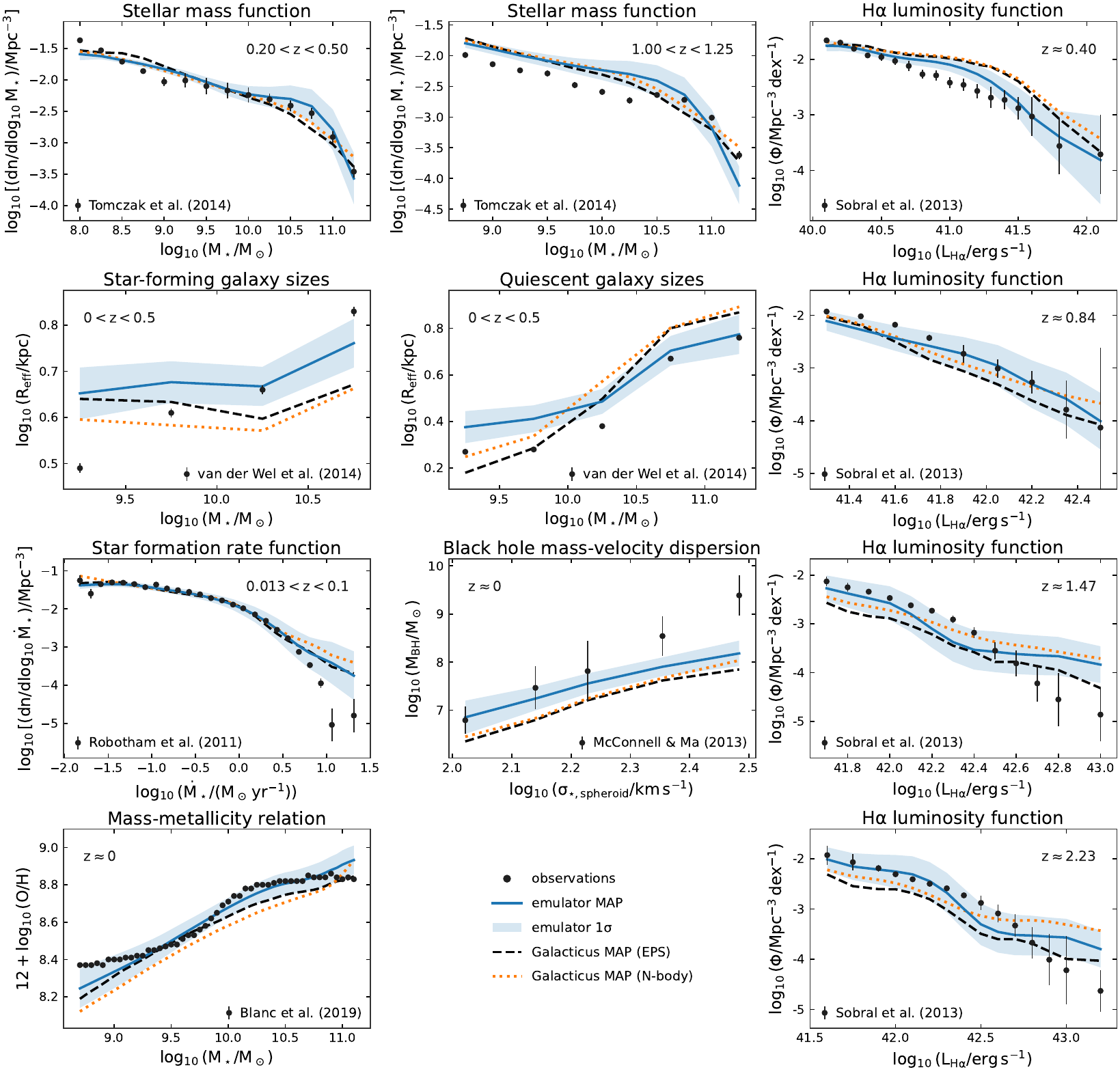}
\caption{
Final joint-calibration predictions for the full calibration-observable set.
The left two columns show the non-\ha{} calibration observables, and the right
column shows the dust-attenuated \ha{} luminosity functions in the four narrow
redshift bins from \citet{2013MNRAS.428.1128S} included in the likelihood.
Black points with error bars show the observational datasets listed in
Table~\ref{tab:calibration-observable-summary}.
The solid blue curves show the emulator prediction at the
maximum-a-posteriori (MAP) parameter vector, with shaded regions indicating
the emulator \(1\sigma\) predictive uncertainty.
The dashed black curves show a direct \galacticus{} evaluation using the
same MAP \galacticus{} parameter values on EPS merger trees; for the \ha{}
luminosity functions, this direct evaluation is post-processed using the MAP
dust-attenuation parameters.
The dotted orange curves show the corresponding direct evaluation on UNIT
\(N\)-body merger trees.
The final calibration provides a simultaneous fit to the stellar mass
functions, size--mass relations, star-formation-rate function, black-hole
scaling relation, mass--metallicity relation, and \ha{} luminosity functions.
The agreement between the emulator predictions and direct \galacticus{} calculations is discussed
in Section~\ref{sec:direct-map-validation}.
}
\label{fig:final-standard-observables}
\end{figure*}

We now turn to the final calibration, in which we fit the full set of
calibration observables described in
Section~\ref{sec:calibration-observable-set}.
The posterior is sampled over the 20 \galacticus{} parameters listed in
Table~\ref{tab:parameters}, together with the five dust-attenuation parameters
defined in Appendix~\ref{app:dust-model-details}.

Figure~\ref{fig:final-standard-observables} shows the maximum-a-posteriori (MAP)
solution from this calibration, together with direct \galacticus{}
calculations with the same parameter values.
We first compare the emulator prediction at the MAP point with the
observational data to assess how well this solution reproduces all of the
calibration observables simultaneously.
The separate comparison between the emulator prediction and the direct
\galacticus{} calculations tests the accuracy of the emulator at the MAP point
and is discussed in Section~\ref{sec:direct-map-validation}.

The calibrated model provides a broadly successful simultaneous description
of the fitted observables, although several systematic residuals remain.
The low-redshift stellar mass function, the star-formation-rate function, and
the gas-phase mass--metallicity relation are especially well reproduced.
The higher-redshift stellar mass function shows larger residuals, with the
model tending to overpredict the abundance of lower-mass galaxies.
A similar trend was reported by \citet{2018MNRAS.475.2936K}, who compared
several semi-analytic models and found that models calibrated to reproduce the
low-redshift stellar mass function commonly overpredict the high-redshift mass
function, particularly at the low-mass end.

The model also predicts a black-hole mass--velocity-dispersion relation that
is too shallow. As a result, the predicted black-hole masses are increasingly
below the observations towards larger velocity dispersions. Fitting to this
observable alone does not remove this behaviour, suggesting that it reflects a
limitation of the adopted model prescription rather than simply a trade-off
introduced by the joint calibration. To determine whether the mismatch arises
primarily from the black-hole masses or the predicted velocity dispersions, we
also compared the direct \galacticus{} MAP calculation with the observed
black-hole--bulge-mass relation of \citet{2013ARA&A..51..511K} and the empirically
inferred black-hole--halo-mass relation of \citet{2023MNRAS.518.2123Z}. At the massive end,
the predicted black-hole masses lie approximately 1 dex below these
relations. This suggests that the discrepancy in the
black-hole mass--velocity-dispersion relation primarily reflects under-massive
black holes in the most massive model galaxies.

The size--mass relations also show systematic residuals, particularly at low
stellar masses, where the MAP model predicts sizes that are too large for both
star-forming and quiescent galaxies. As shown in
Fig.~\ref{fig:combining_constraints}, the observed galaxy sizes lie close to
the lower envelope of the predictions spanned by the training set. Reproducing the observed sizes therefore requires the model to occupy a
relatively narrow region of parameter space, making it difficult to match
the sizes while simultaneously satisfying the constraints from the other
observables.

For the \ha{} luminosity functions shown in the right column of Fig.~\ref{fig:combining_constraints}, the MAP emulator prediction captures the
broad normalisation, shape, and redshift evolution of the observed luminosity
functions across the four narrow redshift bins.
This broad agreement is particularly important for the \romanspace{}  grism
application, for which the abundance of emission-line galaxies is a central
model prediction.

Appendix~\ref{app:calibrated-model} summarises the posterior distribution of
the 25 fitted parameters. Most of the marginalised posteriors retain
substantial support across the regions favoured by their adopted priors.
In contrast, the clearest shift is seen for the disk star-formation frequency
normalisation, \(\nu_{\rm SF,d}\), which is constrained towards the lower
edge of its prior. This preference is already apparent in the diagnostic
calibrations of Fig.~\ref{fig:combining_constraints}: the stellar mass
functions favour low values of \(\nu_{\rm SF,d}\), whereas the
star-formation-rate function alone favours substantially higher values.

In the adopted disk star-formation prescription, the molecular fraction is
set by the \citet{2006ApJ...650..933B} pressure relation, while \(\nu_{\rm SF,d}\) sets the rate at which
molecular gas is converted into stars. For comparison,
\citet{2008AJ....136.2782L} inferred an approximately constant molecular-gas
star-formation efficiency of \(5.25\times10^{-10}\,{\rm yr}^{-1}\) in nearby
spiral galaxies, corresponding to a molecular-gas depletion time of approximately 1.9 Gyr. The MAP value found here, \(1.01\times10^{-10}\,{\rm yr}^{-1}\), instead implies a depletion time of approximately 10 Gyr. The joint
calibration therefore favours substantially slower conversion of molecular
gas into stars than is inferred for nearby spiral galaxies.

This should not necessarily be interpreted as evidence for such long
molecular-gas depletion times in real galaxies. In the calibration,
$\nu_{\rm SF,d}$ is inferred from observables that respond to several coupled
processes, rather than from direct measurements of molecular gas masses and
star-formation rates. Its posterior can therefore shift to offset other aspects
of the model, such as gas supply, feedback, or the partitioning of cold gas into
molecular and atomic phases. Indeed, Fig.~\ref{fig:combining_constraints} shows
that different calibration observables favour different values of
$\nu_{\rm SF,d}$. The low value preferred by the joint calibration should
therefore be interpreted as an effective model parameter within this calibration,
rather than as a direct measurement of molecular-gas depletion times.

The dust-attenuation parameters are also summarised in
Appendix~\ref{app:calibrated-model}. Their posteriors favour a lower
attenuation normalisation and a stronger stellar-mass dependence than the
baseline \gbten{} relation. Because the attenuation model is fitted simultaneously with the
galaxy-formation parameters, the inferred dust relation partly reflects the
intrinsic \ha{} luminosities predicted by \galacticus{}. These dust parameters
should therefore be interpreted within the joint model, rather than as an
independent measurement of the \ha{} attenuation--stellar-mass relation.

\subsection{Direct \galacticus{} Validation At The MAP Point}
\label{sec:direct-map-validation}

Figure~\ref{fig:final-standard-observables} includes two direct \galacticus{}
evaluations at the final MAP parameter vector. The first uses the same set of EPS
merger trees as was used for the emulator training set. This provides a direct
validation of the emulator prediction at the MAP point. Across the calibration observables, the direct EPS \galacticus{} calculation
closely follows the emulator MAP prediction, with residuals generally comparable
to the emulator predictive uncertainty. This is the expected scale for a direct
model evaluation at an emulator-derived optimum, and confirms that the MAP
solution corresponds to a \galacticus{} model that provides a good simultaneous
description of the calibration observables.

The same MAP parameter vector was also evaluated using \(N\)-body merger trees
from one of the UNIT simulations \citep{2019MNRAS.487...48C}. Owing to the large volume of UNIT, we did not run \galacticus{} over the full
simulation. Instead, we randomly selected 0.5 per cent of the merger trees,
corresponding in number density to approximately twelve times the effective
volume of the EPS training runs. The trees are selected randomly from the
full simulation volume rather than from a contiguous subvolume, avoiding
additional sample variance associated with the spatial clustering of haloes. For the \ha{} luminosity functions, this direct \galacticus{}
run was post-processed using the MAP dust-attenuation parameters, matching the
treatment of the EPS calculation.

The resulting UNIT predictions are generally close to those obtained with the
EPS trees. The differences are small for most of the low-redshift observables,
although a noticeable offset is present in the mass--metallicity
relation. Differences become somewhat larger for the higher-redshift
\ha{} luminosity functions. In our EPS calculation, the abundance of the
$z=0$ host haloes is set directly by the adopted Sheth--Tormen halo mass
function. At earlier redshifts, however, the halo population depends
increasingly on how accurately the EPS algorithm reproduces the progenitor
populations of the $z=0$ haloes. Comparisons with $N$-body simulations have
shown that progenitor mass functions and halo assembly histories are sensitive
to the details of the merger-tree construction
\citep[e.g.][]{2008MNRAS.383..546C,2008MNRAS.383..557P}.
The somewhat larger differences seen at higher redshift are therefore not
unexpected.

Overall, these tests show that the good simultaneous fit obtained at the
emulator-derived MAP point is retained when the model is evaluated directly
with \galacticus{}, and that applying the same calibrated parameters to UNIT
\(N\)-body merger trees gives broadly similar predictions. For a future
calibration intended primarily for use with \(N\)-body merger trees, it might
nevertheless be preferable to construct the emulator training set using
\(N\)-body trees. This consideration becomes more important as emulator
uncertainties are reduced, since systematic differences associated with the
merger-tree construction may then become more significant relative to the
emulator uncertainty.

\section{Discussion}
\label{sec:discussion}

\subsection{What Emulator-Assisted Calibration Enables}

A central advantage of emulator-assisted calibration is that, once the
emulator has been trained, the calibration can be repeated at relatively
little computational cost. This makes it practical to perform separate MCMC
analyses using different combinations of observables or different prior assumptions. Comparing the resulting posterior
distributions can reveal which observables constrain particular parameters,
which drive the model towards particular regions of parameter space, and where
different observables favour incompatible parameter values. The examples in
Section~\ref{sec:degeneracy-breaking} illustrate how this can be used to identify degeneracies
and tensions between calibration observables.

This capability becomes increasingly important as the number of varied
parameters grows. In general, converged MCMC exploration becomes more
demanding in higher-dimensional spaces, making repeated calibration with a
computationally expensive galaxy-formation model increasingly impractical.
Fast likelihood evaluations therefore make it feasible to vary a broader set
of \galacticus{} parameters simultaneously. This matters because few
parameters in semi-analytic models have precisely known a priori values.
Holding uncertain parameters fixed makes the model more predictive, but that
predictive power is conditional on assumptions about their values and may
exclude otherwise viable regions of parameter space. Emulator-assisted
calibration makes it practical to relax more of these assumptions and explore
a larger fraction of the model's allowed behaviour.

The emulator also provides a useful way to assess new observations. Rather than
testing only whether an existing calibrated or fiducial model agrees with a new
data set, one can ask whether agreement can be achieved elsewhere within the
allowed model parameter space. This question can be posed either on its own, to
test the flexibility of the model, or jointly with the existing calibration data,
to ask whether the new observable can be accommodated without degrading the
previous fit. An observable that is poorly reproduced by the fiducial model might
become consistent with the data after varying a particular subset of parameters.
Conversely, an observable that cannot be reproduced anywhere in the relevant
region of parameter space provides evidence either of a limitation of the
modelling framework or of tension between observational constraints.

The same approach could also be applied to observables directly relevant for cosmological inference. For example, by constructing training sets from a suite of $N$-body simulations spanning different cosmologies, one could emulate galaxy clustering statistics, such as the galaxy two-point correlation function, as functions of both cosmological and \galacticus{} parameters. Uncertainty in galaxy formation could then be marginalised over when constraining cosmology. Similar approaches have recently been demonstrated using empirical models of the galaxy–halo connection, including subhalo abundance matching and its extension to star-forming galaxies \citep{2026MNRAS.545f2143M,2026arXiv260419449O}. A \galacticus{} emulator would provide a route to propagating uncertainties in a physics-based model of galaxy formation into cosmological constraints.

\subsection{Scope and Limitations}

The calibration presented here should be interpreted within the fixed
modelling choices adopted for this study. The emulator varies continuous
parameter values, but it does not explore the full space of physical
prescriptions available in \galacticus{}. Different choices for feedback,
cooling, dust, environmental processes, merger treatments, or other discrete
model components could introduce different parameters and change how the
resulting parameter space maps onto the observables. The posterior constraints
therefore describe the preferred region of the specified model family, not the
full uncertainty associated with semi-analytic galaxy formation modelling.

The observational likelihood is also simplified. In particular, treating bins
as independent neglects correlations that can arise from shared systematics,
sample variance, calibration choices, and assumptions entering the inference
of galaxy properties. Such correlations can be important when assessing
agreement with a model. For example, a correlated uncertainty may allow an
observable to shift coherently across several bins, whereas treating the same
bins as independent would strongly penalise such a shift. Where available, full observational covariance matrices could instead be incorporated directly into the likelihood.

The finite-sampling uncertainties in the emulator training outputs may likewise be correlated between observable bins. Accounting for these correlations is possible in principle, for example through multi-output Gaussian processes \citep{alvarez2012ftml-kernels} or, for the PCA emulators, by propagating the output covariance through the PCA transformation. Doing so would require an extension of the implementation used here.\footnote{Our emulators use the \texttt{scikit-learn} Gaussian-process implementation, which allows heteroscedastic training uncertainties to be supplied for individual training points, but does not support a full covariance matrix for correlated training errors.} The simpler treatment used here is therefore a practical approximation rather than a statement that such correlations are negligible.

\subsection{Future Work}

A clear next step is to extend the calibration to higher redshift. JWST has revealed unexpectedly abundant bright galaxy candidates at $z \gtrsim 9$, prompting renewed discussion of whether models calibrated primarily to lower-redshift data require changes in star-formation efficiency, burstiness, dust attenuation, or other early-Universe physics \citep[e.g.][]{2023MNRAS.519..843M, 2024ApJ...976L..15C}. Emulator-assisted calibration would make it possible to ask whether a single region of \galacticus{} parameter space can simultaneously accommodate low-redshift constraints and high-redshift JWST measurements, and to identify which physical parameters are driven towards different values by the two regimes.

Future work could also use targeted or sequential training designs. The present emulator is trained over a broad prior volume, which is valuable for global exploration but becomes inefficient once the posterior is localised to a much smaller region of parameter space. Additional \galacticus{} evaluations could instead be targeted towards regions of high posterior probability or large emulator uncertainty, improving the emulator where additional accuracy matters most for the inference. Related iterative strategies have previously been used in emulator-based studies of \textsc{Galform}, where successive waves of model evaluations were concentrated within the regions of parameter space retained by \emph{Bayesian history matching} \citep{2010MNRAS.407.2017B, 2014arXiv1405.4976V, 2017MNRAS.466.2418R}. Such sequential designs may be particularly valuable in high-dimensional parameter spaces, where densely sampling the full prior volume rapidly becomes prohibitively expensive.

A longer-term goal is to extend the framework beyond parameter calibration to
comparisons between alternative physical prescriptions within \galacticus{}.
This would allow the data to constrain not only the values of continuous model
parameters, but also choices between different treatments of the underlying
galaxy-formation physics. Such an extension is more challenging because alternative prescriptions may introduce different parameters, while the Gaussian-process emulators used here are naturally formulated for interpolation over continuous parameter spaces rather than between discrete model choices. Possible approaches include
training separate emulators for different model families or developing
surrogate models that explicitly accommodate categorical model choices.
Although technically more involved, such approaches would begin to address
uncertainty in the physical prescriptions themselves, rather than only
uncertainty in the continuous parameters of a fixed model family.

\section{Conclusions}
\label{sec:conclusions}

We have used Gaussian-process emulators to calibrate a \galacticus{}
semi-analytic galaxy-formation model using a broad set of galaxy observables.
The calibration is motivated by the need for realistic galaxy catalogues to
support the design, interpretation, and analysis of the \romanspace{} Galaxy
Redshift Survey, with emission-line luminosities predicted consistently with
the other properties of the model galaxies.

We constructed a training set of 1024 full \galacticus{} calculations spanning
20 galaxy-formation parameters, and trained Gaussian-process emulators to
predict the observables entering the calibration likelihood. We then used
these emulators to sample the model posterior while accounting for their
predictive uncertainties. The calibration simultaneously includes stellar
mass functions, the star-formation-rate function, size--mass relations, the
black-hole mass--velocity-dispersion relation, the gas-phase mass--metallicity
relation, and dust-attenuated \ha{} luminosity functions, with five additional
parameters describing the dust attenuation model. Finally, we evaluated the
MAP parameter vector directly with \galacticus{} to verify the calibration
obtained using the emulator.

Held-out validation shows that the differences between emulator predictions
and direct \galacticus{} calculations are broadly consistent with the
estimated predictive uncertainties, indicating that these uncertainties are
reasonably well calibrated. We account for the emulator predictive uncertainty
when comparing model predictions with the observational data, allowing the
emulators to be used for posterior inference without neglecting uncertainty
associated with their interpolation of the \galacticus{} calculations.

Combining multiple observables substantially reduces parameter degeneracies
that arise when each observable is considered in isolation. The diagnostic
calibrations also show how different observables can favour different regions
of parameter space. In particular, the weaker stellar feedback favoured by
the galaxy size constraints increases the predicted abundance of galaxies at
fixed stellar mass, causing the stellar mass functions to lie above the
observations. The joint calibration must therefore compromise between the
size and abundance constraints.

The final calibration gives a broadly successful simultaneous description of
the fitted galaxy population, including the abundance and redshift evolution
of the \ha{} emitters that are particularly relevant to the \romanspace{} Galaxy
Redshift Survey. A direct \galacticus{} calculation at the emulator-derived
MAP point gives similar predictions, confirming that the good simultaneous
fit found using the emulator is retained when the model is evaluated directly.
Systematic discrepancies nevertheless remain: the model overpredicts the
abundance of lower-mass galaxies in the higher-redshift stellar mass function,
predicts sizes that are too large at low stellar mass, and produces black
holes that are under-massive in the most massive galaxies. These residuals
provide useful targets for future development of \galacticus{}.

Applying the same calibrated parameters to merger trees extracted from a UNIT
\(N\)-body simulation produces broadly similar predictions to those obtained
with the EPS trees used for emulator training. This supports applying the
calibrated model to \(N\)-body merger trees more generally. Differences
between the EPS- and \(N\)-body-based predictions remain, particularly for
some higher-redshift observables, suggesting that future calibrations may
benefit from constructing the emulator training set directly from \(N\)-body
merger trees, especially if emulator uncertainties are reduced.

The calibrated model provides a practical basis for constructing realistic mock galaxy populations to support the design, interpretation, and analysis of the \romanspace{} Galaxy Redshift Survey. More
generally, our results demonstrate that emulation makes posterior calibration
against many complementary observables practical for galaxy-formation models
that would otherwise be too computationally expensive to explore directly.
The same approach can be applied with different calibration datasets, model
prescriptions, or target applications.

\section*{Acknowledgements}

AR and AB gratefully acknowledge funding from NASA Grant  \#80NSSC24M0021, ``Project Infrastructure for the Roman Galaxy Redshift Survey''. Calculations in this work were performed using the OBS HPC resource operated by Carnegie Science as well as the Resnick High Performance Computing Center, a facility supported by Resnick Sustainability Institute at the California Institute of Technology. This work made use of the following software packages: \href{https://www.astropy.org/}{{Astropy}}
\citep{astropy1, astropy2},
\href{https://emcee.readthedocs.io/en/stable/}{{emcee}}
\citep{ForemanMackey:2013io},
\href{https://github.com/galacticusorg/galacticus}{{Galacticus}}
\citep{2012NewA...17..175B},
\href{https://getdist.readthedocs.io/en/latest/intro.html}{{GetDist}}
\citep{2019arXiv191013970L},
\href{https://www.h5py.org/}{{h5py}}
\citep{h5py},
\href{https://matplotlib.org/}{{Matplotlib}}
\citep{matplotlib},
\href{https://numpy.org/}{{NumPy}}
\citep{numpy},
\href{https://pandas.pydata.org/}{{pandas}}
\citep{pandas-v305, mckinney-proc-scipy-2010},
\href{https://scikit-learn.org/}{{scikit-learn}}
\citep{scikit-learn},
and
\href{https://scipy.org/}{{SciPy}}
\citep{scipy}.

OpenAI's Codex\footnote{\href{https://openai.com/codex/}{https://openai.com/codex/}} was used to assist with coding the emulator and with manuscript preparation. The emulator implementation was reviewed by the authors and validated using the cross-validation and held-out-\galacticus{} tests described in Section~\ref{sec:emulator-validation} and Appendix~\ref{app:emulator-validation-summary}. All AI-assisted output was reviewed, edited, and verified by the authors, who take full responsibility for the content of the manuscript.

\bibliographystyle{aasjournal}
\bibliography{bibliography}

\appendix

\section{Emission-Line Dust Model Details}
\label{app:dust-model-details}

This appendix describes the dust attenuation applied to the emission-line luminosities used in the \ha{} luminosity-function calibration. The intrinsic line luminosities are calculated within \galacticus{}, with the emission-line model described in detail by Weerasooriya et al. (in preparation). For star-forming regions, the model uses pre-computed \textsc{Cloudy} photoionization calculations, as described in Section~\ref{sec:dust-postprocessing}. AGN emission lines are calculated using a separate \textsc{Cloudy} tabulation with spectra appropriate for AGN, with the luminosity of the ionizing source determined from the black hole accretion rate and the \galacticus{} accretion-disk model. We sum the contributions from star-forming regions in the disk and spheroid together with those from the AGN to obtain the intrinsic line luminosities. These are attenuated using the model below before constructing luminosity functions in the \citet{2013MNRAS.428.1128S} bins.

The baseline attenuation model is the local \ha{} attenuation--stellar-mass
relation of \gbten{}. Defining
\begin{equation}
  X = \log_{10}\left(\frac{M_\star}{10^{10}\msun}\right),
\end{equation}
the baseline attenuation is
\begin{equation}
  A_{\Hamath}^{\rm GB10}
  =
  0.91 + 0.77X + 0.11X^2 - 0.09X^3 .
  \label{eq:gb10-attenuation}
\end{equation}
To avoid extrapolating this empirical relation far beyond the stellar-mass range
over which it was calibrated, we clip the stellar mass entering
Equation~\ref{eq:gb10-attenuation} to the range
$10^8 < M_\star/\msun < 10^{11}$ before evaluating $X$.

We allow the attenuation relation to depart from this baseline according to
\begin{equation}
  \bar{A}_{\Hamath}(M_\star,z)
  =
  A_{\Hamath}^{\rm GB10}(M_\star)
  + \delta_0
  + \delta_M X
  + \delta_z u
  + \delta_{Mz}Xu ,
  \label{eq:halpha-dust-model}
\end{equation}
where
\begin{equation}
  u = \ln \left( \frac{1+z}{1+z_{\rm piv}} \right),
\end{equation}
and we adopt $z_{\rm piv}=1$. The parameters $\delta_0$ and $\delta_M$
therefore allow departures from the normalisation and stellar-mass dependence
of the \gbten{} relation at the pivot redshift, while $\delta_z$ and
$\delta_{Mz}$ allow these departures to evolve with redshift.

For an individual galaxy, the attenuation is described by
\begin{equation}
  A_{\Hamath}(M_\star,z)
  =
  \bar{A}_{\Hamath}(M_\star,z) + \epsilon_A,
  \qquad
  \epsilon_A \sim \mathcal{N}(0,\sigma_A^2),
\end{equation}
with negative values truncated at $A_{\Hamath}=0$.

The five dust parameters sampled in the final calibration are
$\delta_0$, $\delta_M$, $\delta_z$, $\delta_{Mz}$, and $\sigma_A$.
We adopt independent normal priors with mean zero and standard deviation 0.5
for the four shift parameters. For $\sigma_A$, we adopt a normal prior with
mean 0.25 and standard deviation 0.1, truncated at $\sigma_A=0$. These priors
allow departures from the local \gbten{} relation while retaining it as the
reference model for nebular attenuation.

The attenuated \ha{} luminosity is
\begin{equation}
  \log_{10} L_{\Hamath}^{\rm obs}
  =
  \log_{10} L_{\Hamath}^{\rm int}
  - 0.4 \, A_{\Hamath}.
\end{equation}
Rather than drawing a stochastic attenuation for each galaxy when
$\sigma_A>0$, we compute the expected contribution of each galaxy to each
luminosity bin by integrating over the attenuation distribution, including the
truncation at $A_{\Hamath}=0$. This avoids introducing Monte Carlo noise into the luminosity functions while retaining the redistribution of galaxies between luminosity bins caused by attenuation scatter.

For each of the 1024 \galacticus{} training points, we independently draw one
set of dust parameters from these priors, giving 1024 training points
distributed over the combined 25-dimensional parameter space. Because the dust
attenuation is applied in post-processing, additional dust realisations can be
generated at negligible cost compared with additional \galacticus{} runs. We tested training the emulator with multiple dust realisations per
\galacticus{} run, but found only modest improvements in held-out predictive
accuracy. We therefore use one dust realisation per run for the fiducial
emulator, keeping the training set smaller and reducing the computational cost
of emulator training and evaluation. The \ha{} luminosity-function emulator consequently has 25 input
dimensions: the 20 \galacticus{} parameters and the five dust parameters.

% Dust-draw comparison provenance:
% runs/campaigns/sobol_mass_function_emissionlines_dust_simpleSizes_freeYield_20p_512_reduced/
% cross_validation/emission_line_lfs/holdout_80_20/
% halpha_sobral_1_vs_2_vs_4dustDraws_comparison/
% halpha_sobral_1_vs_2_vs_4dustDraws_summary.csv

\section{Observable And Emulator Details}
\label{app:observable-details}

\subsection{Held-Out Emulator Validation Summary}
\label{app:emulator-validation-summary}

The RMSE and $R^2$ statistics defined in
Section~\ref{sec:emulator-validation} quantify the agreement between the
emulator mean and the direct \galacticus{} calculation. They are useful
descriptive measures of predictive accuracy, and are used in
Figures~\ref{fig:example_predicted_vs_heldout} and \ref{fig:training_size}.
However, neither statistic isolates the interpolation error of the emulator.
Each held-out \galacticus{} prediction is measured from a finite simulated
galaxy population and is therefore itself a noisy estimate of the underlying
model prediction.

For a given observable bin, let $g_n$ be the value measured from held-out
\galacticus{} calculation $n$, and let $f_n$ denote the corresponding infinite-sampling model prediction, i.e. the expectation that would be approached as the number of simulated haloes becomes arbitrarily large. We write
\begin{equation}
  g_n = f_n + \epsilon_n,
  \qquad
  {\rm Var}(\epsilon_n) = \sigma_{{\rm sim},n}^2,
\end{equation}
where $\sigma_{{\rm sim},n}$ is the estimated finite-sampling uncertainty of
the held-out calculation. This is the same type of uncertainty that is
supplied to the GP as a training-output uncertainty when building the emulators.

Let $m_n$ be the emulator mean at the held-out parameter point, and let
$\sigma_{{\rm emu},n}$ be its predictive uncertainty for $f_n$, the
infinite-sampling model prediction defined above. Treating the emulator
predictive uncertainty and finite-sampling uncertainty as independent, the
expected variance of the residual, $m_n - g_n$, is
\begin{equation}
  s_n^2
  =
  \sigma_{{\rm emu},n}^2
  +
  \sigma_{{\rm sim},n}^2.
\end{equation}
We therefore define the standardised residual
\begin{equation}
  z_n = \frac{m_n-g_n}{s_n}.
\end{equation}
For each observable bin, we summarise these residuals using
\begin{equation}
  {\rm RMS}(z)
  =
  \left[
    \frac{1}{N_{\rm val}}
    \sum_n z_n^2
  \right]^{1/2},
\end{equation}
and the coverage fractions
\begin{equation}
  f_{k\sigma}
  =
  \frac{1}{N_{\rm val}}
  \sum_n
  \mathbb{I}\!\left(\lvert z_n\rvert \leq k\right),
  \qquad k\in\{1,2\},
\end{equation}
where $\mathbb{I}$ is the indicator function. If the combined uncertainties
are correctly calibrated and the standardised residuals are approximately
Gaussian, we expect ${\rm RMS}(z)\approx1$, $f_{1\sigma}\approx0.68$, and
$f_{2\sigma}\approx0.95$.

Table~\ref{tab:emulator-validation-summary} gives the median value of each
diagnostic over the bins of an observable. The raw RMSE values measure the
difference between the emulator mean and the finite-volume held-out
\galacticus{} catalogue prediction, rather than the emulator interpolation error
alone. They therefore include noise from the direct held-out calculation itself,
as well as uncertainty in the emulator prediction.

This distinction is especially important for the \ha{} luminosity functions. For
a survey covering a fixed area, the comoving volume probed by the observational
samples generally increases towards higher redshift, allowing the observed
luminosity functions to extend to lower number densities. By contrast, every
\galacticus{} calculation used here has the same effective comoving volume. The
higher-redshift luminosity-function bins therefore contain fewer simulated
galaxies and have larger finite-sampling uncertainties, making the direct
held-out predictions noisier at higher redshift.

After normalising the residuals by the combination of emulator predictive
uncertainty and held-out finite-sampling uncertainty, the \ha{}
luminosity-function emulators have ${\rm RMS}(z)$ values and coverage fractions
that are reasonably close to the Gaussian expectation at all four redshifts. The
larger raw RMSE values at higher redshift should therefore not be interpreted as
a comparable deterioration in the emulator accuracy for the underlying
infinite-sampling model prediction.

\begin{table*}
\centering
\caption{
Five-fold held-out validation summary for the calibration-observable
emulators trained on the full 1024-run campaign.
RMSE and $R^2$ are defined in Section~\ref{sec:emulator-validation}.
For each held-out prediction, the standardised residual $z_n$ is normalised
by the combined emulator predictive uncertainty and finite-sampling uncertainty
of the held-out \galacticus{} calculation, as defined in
Appendix~\ref{app:emulator-validation-summary}.
The entries are medians over the bins of each observable.
$f_{1\sigma}$ and $f_{2\sigma}$ are the fractions of held-out predictions
with $\lvert z_n\rvert\leq1$ and $\lvert z_n\rvert\leq2$, respectively.
}
\label{tab:emulator-validation-summary}
\begin{tabular}{@{}p{0.3\textwidth}p{0.14\textwidth}rrrrr@{}}
\toprule
Calibration observable
  & Redshift
  & RMSE
  & $R^2$
  & ${\rm RMS}(z)$
  & $f_{1\sigma}$
  & $f_{2\sigma}$ \\
\midrule

\multirow{2}{0.3\textwidth}{Stellar mass function}
  & $0.20<z<0.50$ & 0.11 & 0.92 & 1.17 & 0.69 & 0.94 \\
  & $1.00<z<1.25$ & 0.24 & 0.92 & 1.14 & 0.66 & 0.92 \\
\addlinespace[0.5em]

Star-formation-rate function
  & $0.013<z<0.1$ & 0.11 & 0.88 & 1.08 & 0.70 & 0.94 \\
\addlinespace[0.5em]

Size--mass relation (star-forming)
  & $0<z<0.5$ & 0.05 & 0.90 & 1.14 & 0.69 & 0.93 \\

Size--mass relation (quiescent)
  & $0<z<0.5$ & 0.06 & 0.84 & 1.13 & 0.70 & 0.93 \\
\addlinespace[0.5em]

Black-hole--velocity-dispersion relation
  & $z\approx0$ & 0.32 & 0.79 & 1.40 & 0.64 & 0.90 \\
\addlinespace[0.5em]

Gas-phase mass--metallicity relation
  & $z\approx0$ & 0.08 & 0.89 & 1.18 & 0.74 & 0.94 \\
\addlinespace[0.5em]

\multirow{4}{0.3\textwidth}{\ha{} luminosity function}
  & $z=0.40$ & 0.14 & 0.85 & 1.11 & 0.67 & 0.94 \\
  & $z=0.84$ & 0.60 & 0.78 & 1.11 & 0.67 & 0.92 \\
  & $z=1.47$ & 0.74 & 0.78 & 1.05 & 0.72 & 0.94 \\
  & $z=2.23$ & 0.79 & 0.78 & 1.05 & 0.73 & 0.93 \\
\bottomrule
\end{tabular}
\end{table*}
% Provenance:
% The non-H-alpha rows were computed from
% runs/campaigns/sobol_1024_20p_simpleSizes_moreHalos_reduced/cross_validation/table3_kfold_5/training_size_cv_metrics.csv
% using rows with emulator_type=pca, subset_size=1024, validation_mode=kfold,
% bin != all, and uncertainty_denominator=emulator_plus_heldout.
% Each table entry is the median over bins of that observable.
%
% The H-alpha rows use
% runs/campaigns/sobol_1024_20p_simpleSizes_moreHalos_reduced/cross_validation/sidecar_lf_kfold_5/halpha_sobral_1dustdraw_pca99/halpha_sobral_1dustdraw_pca99_predictions.csv
% with held-out finite-sampling uncertainties read from each corresponding
% evaluations/<evaluation_id>/emission_line_dust/emission_line_dust_lf_emulator_table.csv
% column:
%   <halpha_sobral_z*_bin*>_shot_noise_log10_std_conservative.
% Nonfinite held-out H-alpha uncertainties, which occur for zero-count bins,
% were treated as the conservative 2 dex cap, then all held-out uncertainties
% were clipped to [1e-4, 2] dex before computing
% z_i = (y_pred - y_true) / sqrt(y_pred_std^2 + sigma_holdout^2).
% H-alpha RMSE, R^2, RMS(z), f_1sigma, and f_2sigma were computed per redshift
% and bin, then the table entries were taken as medians over bins.

\subsection{Treatment of floored and missing emulator training targets}
\label{app:missing-values}

The transformations described in
Section~\ref{sec:observable-space-transformations} require a finite emulator
target for every training model and observable bin. This appendix specifies
how model predictions that did not satisfy this requirement were treated.
These operations apply only to the model predictions used to train the emulators; no values in the observational calibration data are replaced or modified.

\subsubsection{Number-density observables}

For the stellar-mass and star-formation-rate functions, predictions at or below
a minimum number density were assigned
\[
  \log_{10}\left(
    \frac{\Phi_{\rm floor}}
         {{\rm Mpc}^{-3}\,{\rm dex}^{-1}}
  \right)
  = -6.64.
\]
For the dust-attenuated \ha{} luminosity functions, the corresponding floor was
\[
  \log_{10}\left(
    \frac{\Phi_{{\rm floor},{\rm H}\alpha}}
         {{\rm Mpc}^{-3}\,{\rm dex}^{-1}}
  \right)
  = -8.
\]
Floored stellar-mass- and star-formation-rate-function predictions were
assigned a training uncertainty of \(0.5\) dex, while floored \ha{} luminosity
functions were assigned a training uncertainty of \(2\) dex. These enlarged uncertainties reduce the influence of floored predictions on the emulator training.

Both floors lie well below the observational data entering the final
calibration. The smallest observed number density is a factor of \(40\) above
the corresponding floor, for the star-formation-rate function, while the
stellar-mass and \ha{} luminosity functions remain at least three orders of
magnitude above their floors.

\subsubsection{Undefined binned relations}

For binned relations, the predicted ordinate is undefined if a model produces
no galaxies satisfying the selection for a particular bin. Such entries must
nevertheless be assigned finite values because the PCA and
Gaussian-process training require a complete data vector for every training
model. This affected the emulated ordinates
\(\log_{10}R_{\rm eff}\), \(\log_{10}M_{\rm BH}\), and
\(12+\log_{10}({\rm O/H})\).

Each undefined entry was replaced by the median of the valid training-set predictions in the same bin, with the replacement therefore determined independently for each bin. Each replaced entry was assigned a training uncertainty of 5 in the units of the emulated ordinate, so that it contributes little information to the emulator training.

\subsubsection{Incidence in the production training set}

Table~\ref{tab:missing-training-targets} reports the incidence of flooring and
median replacement among the \(1024\) production training models. A bin is
listed as affected if at least one training model required treatment. The
affected fraction is the fraction of all training-model--bin combinations for
the corresponding table row that required treatment.

\begin{table*}
\centering
\caption{
Floored or median-replaced predictions in the production emulator training
set. Each observable has 1024 training predictions per bin. The affected
region gives the coordinates of bins in which at least one prediction required
treatment, while the affected-points fraction is calculated over all training
models and bins for the corresponding observable.
}
\label{tab:missing-training-targets}
\begin{tabular}{@{}p{0.28\textwidth}p{0.15\textwidth}p{0.1\textwidth}p{0.24\textwidth}r@{}}
\toprule
Calibration observable
  & Treatment
  & Affected bins
  & Affected region
  & Affected fraction\\
\midrule

Black-hole--velocity-dispersion relation
  & Median replacement
  & \(2/5\)
  & \(\sigma_\star\geq226\,{\rm km\,s^{-1}}\)
  & \(8.0\%\) \\
\addlinespace[0.5em]

\multirow{2}{0.27\textwidth}{Stellar mass function}
  & \multirow{2}{0.15\textwidth}{Floor}
  & \(5/14\)
  & \(M_\star\geq10^{10.25}\,\msun\), \(0.2<z<0.5\)
  & \(3.9\%\) \\
  & & \(6/11\)
  & \(M_\star\geq10^{10.00}\,\msun\), \(1<z<1.25\)
  & \(8.9\%\) \\
\addlinespace[0.5em]

Gas-phase mass--metallicity relation
  & Median replacement
  & \(18/49\)
  & \(M_\star\geq10^{10.25}\,\msun\)
  & \(1.1\%\) \\
\addlinespace[0.5em]

Star-formation-rate function
  & Floor
  & \(8/25\)
  & \(\dot M_\star\gtrsim2.0\,\mathrm{M_\odot\,yr^{-1}}\)
  & \(3.0\%\) \\
\addlinespace[0.5em]

Size--mass relation (star-forming)
  & Median replacement
  & \(1/4\)
  & \(M_\star=10^{10.75}\,\msun\)
  & \(0.1\%\) \\

Size--mass relation (quiescent)
  & Median replacement
  & \(2/5\)
  & \(M_\star\geq10^{10.75}\,\msun\)
  & \(1.4\%\) \\
\addlinespace[0.5em]

\multirow{4}{0.27\textwidth}{\ha{} luminosity function}
  & \multirow{4}{0.15\textwidth}{Floor}
  & \(5/18\)
  & \(L_{\mathrm{H}\alpha}\geq10^{41.4}\,{\rm erg\,s^{-1}}\), \(z=0.40\)
  & \(0.7\%\) \\
  & & \(7/9\)
  & \(L_{\mathrm{H}\alpha}\geq10^{41.6}\,{\rm erg\,s^{-1}}\), \(z=0.84\)
  & \(9.9\%\) \\
  & & \(13/13\)
  & All bins, \(z=1.47\)
  & \(26.0\%\) \\
  & & \(15/15\)
  & All bins, \(z=2.23\)
  & \(30.0\%\) \\
\bottomrule
\end{tabular}
\end{table*}

For the standard observables, treated entries occur predominantly in bins
selecting massive galaxies or high star-formation rates, where some training
models contain few or no qualifying galaxies. The incidence is substantially
larger for the two highest-redshift \ha{} luminosity functions. At these
redshifts, the fixed effective volume of each \galacticus{} calculation
contains fewer galaxies in many of the luminosity bins probed by the
observational data. In addition, some regions of the dust-model parameter
space attenuate the \ha{} emission sufficiently strongly that a model produces
no emitters in a bin, including in some relatively faint luminosity bins.

Both treatments predominantly affect regions of parameter space that produce
galaxy populations inconsistent with the calibration data. A number-density
prediction below the adopted floor is already far below the corresponding
observed abundance. Similarly, an undefined binned relation usually occurs
because a model produces no galaxies satisfying the relevant selection,
typically in a massive or otherwise sparsely populated part of the galaxy
population. Such models will generally also be disfavoured by the
number-density observables that constrain the abundance of these galaxies.
The precise treatment of these predictions is therefore less important than
the accuracy of the emulator in regions of parameter space that provide
plausible fits to the calibration data.

\subsection{Covariance And Uncertainty Treatment}
\label{app:covariance-treatment}

The observable vectors measured from the finite galaxy populations in each
\galacticus{} training calculation have finite-sampling uncertainties. These
uncertainties are estimated separately for each training model and observable
bin and are included when training the Gaussian-process emulators. Here we
describe how the per-bin training uncertainties are propagated through the
standardisation and PCA transformations and how the resulting GP predictive
uncertainties are reconstructed in observable space.

For training model \(n\), let \(y_{nb}\) denote the prediction in observable
bin \(b\), after applying the observable-space transformation described in
Section~\ref{sec:observable-space-transformations}, and let
\(\sigma_{nb}^2\) denote the estimated finite-sampling variance of that
prediction in the same transformed space. We retain only the diagonal per-bin
variances, neglecting correlations in the finite-sampling fluctuations between
different observable bins.

For the PCA emulators, each observable bin is first standardised using its
training-set mean, \(\mu_b\), and standard deviation, \(s_b\), as described in
Section~\ref{sec:pca-compression}:
\begin{equation}
  z_{nb}
  =
  \frac{y_{nb}-\mu_b}{s_b}.
\end{equation}
The corresponding finite-sampling variance is
\begin{equation}
  {\rm Var}(z_{nb})
  =
  \frac{\sigma_{nb}^2}{s_b^2}.
\end{equation}

The coefficient of training model \(n\) along PCA component \(k\) is
\begin{equation}
  a_{nk}
  =
  \sum_b e_{kb}z_{nb},
\end{equation}
where \(e_{kb}\) is the \(b\)th element of the \(k\)th PCA component. Under the
diagonal approximation for the per-bin training covariance, the
finite-sampling variance of this coefficient is
\begin{equation}
  \sigma_{a,nk}^2
  =
  \sum_b e_{kb}^2
  \frac{\sigma_{nb}^2}{s_b^2}.
  \label{eq:pca-training-variance}
\end{equation}
Although the PCA transformation can induce covariance between the
finite-sampling uncertainties of different PCA coefficients, these
cross-component covariances are not retained; each coefficient is modelled
with an independent scalar GP.

Before fitting its GP, each retained PCA coefficient is itself standardised:
\begin{equation}
  \widetilde{a}_{nk}
  =
  \frac{a_{nk}-\bar{a}_k}{q_k},
\end{equation}
where \(\bar{a}_k\) and \(q_k\) are the training-set mean and standard
deviation of coefficient \(k\). The training-output variance supplied to the
GP for this training point is therefore
\begin{equation}
  \sigma_{\widetilde{a},nk}^2
  =
  \frac{\sigma_{a,nk}^2}{q_k^2}.
\end{equation}
These variances are added to the diagonal of the covariance matrix obtained by
evaluating the GP kernel at the training points. The resulting training-noise
term is heteroscedastic: it can differ between training models and between PCA
coefficients.

For the bin-by-bin emulators used in the validation comparisons, the same
procedure is applied without the PCA transformation. The finite-sampling
variance in each bin is propagated through the standardisation of the
corresponding scalar GP target and added to the diagonal of its training
covariance matrix.

At a new point in parameter space, let
\(\sigma_{a,k}^2(\boldsymbol{\theta})\) denote the predictive variance from the
GP for PCA coefficient \(k\), after reversing the standardisation of that
coefficient. Treating the retained PCA-coefficient predictions as independent,
the marginal emulator variance in observable bin \(b\) is
\begin{equation}
  \sigma_{{\rm emu},b}^2(\boldsymbol{\theta})
  =
  s_b^2
  \sum_{k=1}^{K}
  e_{kb}^2
  \sigma_{a,k}^2(\boldsymbol{\theta}).
  \label{eq:pca-predictive-variance}
\end{equation}
The PCA reconstruction also induces covariance between different observable
bins, but only the marginal variance of each bin is retained. These diagonal
emulator variances are used in the calibration likelihood described in
Section~\ref{sec:likelihood}.

The marginal predictive variances are therefore obtained by propagating the
uncertainties of the retained PCA coefficients through the PCA reconstruction.
The adequacy of the 99 per cent variance threshold, including any error
introduced by the compression, is assessed empirically through the held-out
validation tests in Section~\ref{sec:emulator-validation}.

\section{Calibrated Model Posterior}
\label{app:calibrated-model}

Table~\ref{tab:final-calibrated-parameters} summarises the MAP point and
marginalised posterior distribution for the final calibration. In addition
to the parameter values themselves, we report their locations within the
adopted priors, making it easier to identify parameters for which the
calibration favours values in the tails of their prior distributions. As
discussed in Section~\ref{sec:varied-parameters}, the priors are generally
centred on values used in previous \galacticus{} configurations or suggested
by earlier calibration experiments, rather than representing precise prior
constraints on the underlying physical parameters.

% Generated by paper/figure_scripts/generate_final_calibrated_parameters_table.py
% created_utc: 2026-08-20T16:40:25.461395+00:00
% command: python paper/figure_scripts/generate_final_calibrated_parameters_table.py --output runs/campaigns/sobol_1024_20p_simpleSizes_moreHalos_reduced/automatedPipeline_transformedParams_finalPaper_definitive/MCMCs_production_from_exploratory_MAP/standard_observables_plus_emission_line_lfs/all_standard_observables_plus_halpha_sobral_1dustdraw_pca99_sobralLogErr/paperFigures/final_calibrated_parameters_table.tex
% cwd: repository root
% source_hdf5: runs/campaigns/sobol_1024_20p_simpleSizes_moreHalos_reduced/automatedPipeline_transformedParams_finalPaper_definitive/MCMCs_production_from_exploratory_MAP/standard_observables_plus_emission_line_lfs/all_standard_observables_plus_halpha_sobral_1dustdraw_pca99_sobralLogErr/mcmc_all_standard_observables_plus_halpha_sobral_1dustdraw_pca99_sobralLogErr_mcmc_results.hdf5
% git_head: e67c845484ddf4fc3262dd3991032201fe2b11d9

\providecommand{\paramblockspace}{\addlinespace[1.4ex]}

\begin{table*}
\centering
\scriptsize
\setlength{\tabcolsep}{4.2pt}
\renewcommand{\arraystretch}{1.10}
\caption{
Posterior summary for the final calibrated model.
The columns \(p_{16}\), \(p_{50}\), and \(p_{84}\) give the 16th, 50th, and
84th percentiles of the marginalised posterior.
The \(q\) columns give the corresponding locations in the adopted prior,
expressed as percentiles on a 0--100 scale.
Units for dimensional parameters are shown in the parameter column.
}
\label{tab:final-calibrated-parameters}
\begin{tabular}{@{}l @{\hspace{2.1em}} r @{\hspace{2.5em}} rrr @{\hspace{3.1em}} rrrr@{}}
\toprule
Parameter & MAP & \multicolumn{3}{c}{Posterior} & \multicolumn{4}{c}{Prior percentile} \\
\cmidrule(l{1.0em}r{2.0em}){3-5}
\cmidrule(l{0.0em}r{0.0em}){6-9}
& & \(p_{16}\) & \(p_{50}\) & \(p_{84}\) & \(q_{\rm MAP}\) & \(q_{16}\) & \(q_{50}\) & \(q_{84}\) \\
\midrule
\multicolumn{9}{@{}l}{\normalfont Star formation and stellar populations} \\
\(\nu_{\rm SF,d}/\mathrm{Gyr}^{-1}\) & \(0.101\) & \(0.102\) & \(0.107\) & \(0.119\) & 0.1 & 0.3 & 1.0 & 2.5 \\
\(\epsilon_{\star,{\rm sph}}\) & \(0.0839\) & \(0.0860\) & \(0.108\) & \(0.135\) & 77.0 & 77.3 & 79.9 & 82.3 \\
\(y_Z\) & \(0.0362\) & \(0.0328\) & \(0.0355\) & \(0.0382\) & 95.9 & 90.6 & 94.9 & 98.2 \\
\paramblockspace

\multicolumn{9}{@{}l}{\normalfont Stellar feedback} \\
\(V_{\rm out,d}/\mathrm{km\,s^{-1}}\) & \(170\) & \(155\) & \(167\) & \(183\) & 65.2 & 57.6 & 63.6 & 71.3 \\
\(\alpha_{\rm out,d}\) & \(2.55\) & \(2.37\) & \(2.57\) & \(2.73\) & 63.7 & 59.1 & 64.2 & 68.3 \\
\(V_{\rm out,sph}/\mathrm{km\,s^{-1}}\) & \(10.8\) & \(10.5\) & \(11.9\) & \(14.6\) & 1.0 & 0.7 & 2.8 & 6.9 \\
\paramblockspace

\multicolumn{9}{@{}l}{\normalfont Gas cooling and recycling} \\
\(f_{\rm cool}\) & \(0.436\) & \(0.395\) & \(0.447\) & \(0.526\) & 54.4 & 49.6 & 55.7 & 63.8 \\
\(r_{\rm core}/R_{\rm vir}\) & \(0.303\) & \(0.258\) & \(0.330\) & \(0.456\) & 56.4 & 49.2 & 60.2 & 74.4 \\
\(\gamma_{\rm reinc}\) & \(1.64\) & \(1.64\) & \(2.19\) & \(2.82\) & 1.2 & 1.2 & 4.8 & 12.6 \\
\(\delta_{\rm reinc,1}\) & \(-0.127\) & \(-0.935\) & \(-0.450\) & \(-0.0224\) & 31.2 & 17.8 & 25.8 & 33.0 \\
\(\delta_{\rm reinc,2}\) & \(3.60\) & \(2.95\) & \(3.81\) & \(4.50\) & 76.7 & 65.9 & 80.2 & 91.6 \\
\paramblockspace

\multicolumn{9}{@{}l}{\normalfont Mergers and disk instabilities} \\
\(f_{\rm major}\) & \(0.467\) & \(0.344\) & \(0.422\) & \(0.476\) & 97.4 & 79.7 & 92.7 & 98.2 \\
\(f_{\rm orbit}\) & \(1.52\) & \(1.48\) & \(1.94\) & \(2.62\) & 80.1 & 78.7 & 90.9 & 97.6 \\
\(f_{j,{\rm sph}}^{\rm bar}\) & \(0.177\) & \(0.105\) & \(0.157\) & \(0.236\) & 40.3 & 9.6 & 31.3 & 62.8 \\
\paramblockspace

\multicolumn{9}{@{}l}{\normalfont Supermassive black holes} \\
\(M_{\rm BH,seed}/(10^4\mathrm{M_\odot})\) & \(7.17\) & \(1.68\) & \(4.76\) & \(8.21\) & 96.2 & 76.6 & 91.1 & 97.8 \\
\(\epsilon_{\rm BH,radio}\) & \(0.183\) & \(0.171\) & \(0.268\) & \(0.419\) & 16.2 & 13.0 & 41.4 & 75.4 \\
\(\epsilon_{\rm BH,wind}\) & \(0.00321\) & \(0.00150\) & \(0.00404\) & \(0.00937\) & 61.7 & 31.6 & 70.3 & 92.2 \\
\(\alpha_{\rm Bondi,sph}\) & \(0.0524\) & \(0.149\) & \(0.442\) & \(1.26\) & 0.1 & 3.1 & 11.0 & 25.5 \\
\(\alpha_{\rm Bondi,hot}\) & \(3.29\) & \(3.26\) & \(15.3\) & \(49.2\) & 41.0 & 40.8 & 73.7 & 92.7 \\
\(x_{\rm thin,max}\) & \(21.8\) & \(16.9\) & \(23.1\) & \(27.9\) & 95.8 & 92.1 & 96.6 & 99.1 \\
\paramblockspace

\multicolumn{9}{@{}l}{\normalfont Dust attenuation} \\
\(\delta_0\) & \(-0.738\) & \(-0.886\) & \(-0.688\) & \(-0.279\) & 7.0 & 3.8 & 8.4 & 28.9 \\
\(\delta_M\) & \(0.795\) & \(0.0575\) & \(0.711\) & \(1.07\) & 94.4 & 54.6 & 92.2 & 98.4 \\
\(\delta_z\) & \(0.518\) & \(-0.525\) & \(0.146\) & \(0.605\) & 85.0 & 14.7 & 61.5 & 88.7 \\
\(\delta_{Mz}\) & \(0.0474\) & \(-0.155\) & \(0.433\) & \(1.15\) & 53.8 & 37.9 & 80.7 & 98.9 \\
\(\sigma_A\) & \(0.258\) & \(0.208\) & \(0.290\) & \(0.375\) & 53.0 & 33.5 & 65.5 & 89.3 \\
\bottomrule
\end{tabular}
\end{table*}

\begin{figure*}
\centering
\includegraphics[width=\textwidth]{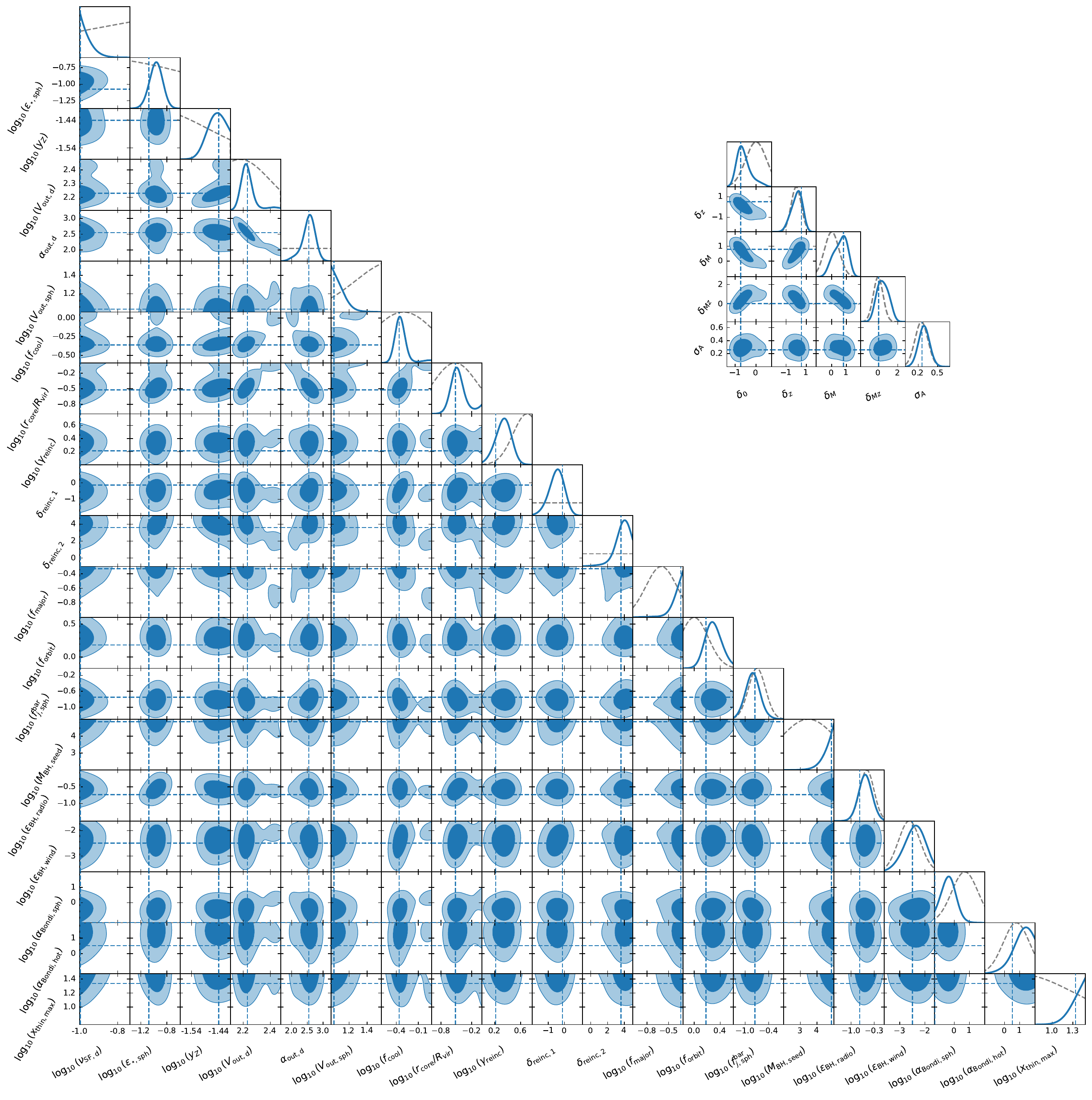}
\caption{
Marginalised posterior distributions for the 20 \galacticus{} parameters in
the final joint calibration.
The inset shows the corresponding corner plot for the five dust-attenuation
parameters.
Diagonal panels show the one-dimensional marginalised posteriors, with dashed
grey curves showing the adopted priors.
Dashed blue vertical and horizontal lines mark the maximum-a-posteriori
parameter values.
Off-diagonal panels show the corresponding two-dimensional marginalised
posterior distributions, with 68 and 95 per cent credible regions.
}
\label{fig:final-posterior-corner}
\end{figure*}

\end{document}